\documentclass[pra,twocolumn,preprintnumbers,amsmath,amssymb,floatfix,longbibliography,superscriptaddress]{revtex4-1}
\usepackage{graphicx}
\usepackage{graphics}
\usepackage{psfrag}
\usepackage{hyperref}
\usepackage{multirow}
\usepackage[sort&compress]{natbib}
\usepackage{amssymb}
\usepackage{amsmath}
\usepackage{amsthm}
\usepackage{bbold}
\usepackage{bbm}
\usepackage{enumerate}
\usepackage{textcomp}

\newcommand{\vphi}{\varphi}

\newcommand{\vare}{\varepsilon}

\newcommand{\rmi}{{\rm i}}

\usepackage{xcolor}

\begin{document}

\hypersetup{pdftitle={title}}
\title{Quantum Simulation of Hyperbolic Space with Circuit Quantum Electrodynamics: From Graphs to Geometry}

\author{Igor Boettcher}
\email{iboettch@umd.edu}
\affiliation{Joint Quantum Institute, University of Maryland, College Park, MD 20742, USA}

\author{Przemyslaw Bienias}
\affiliation{Joint Quantum Institute, University of Maryland, College Park, MD 20742, USA}
\affiliation{Joint Center for Quantum Information and Computer Science, NIST/University of Maryland, College Park, Maryland 20742, USA}

\author{Ron Belyansky}
\affiliation{Joint Quantum Institute, University of Maryland, College Park, MD 20742, USA}

\author{Alicia J. Koll\'{a}r}
\affiliation{Joint Quantum Institute, University of Maryland, College Park, MD 20742, USA}

\author{Alexey V. Gorshkov}
\affiliation{Joint Quantum Institute, University of Maryland, College Park, MD 20742, USA}
\affiliation{Joint Center for Quantum Information and Computer Science, NIST/University of Maryland, College Park, Maryland 20742, USA}

\begin{abstract}
We show how quantum many-body systems on hyperbolic lattices with nearest-neighbor hopping and local interactions can be mapped onto quantum field theories in continuous negatively curved space. The underlying lattices have recently been realized experimentally with superconducting resonators and therefore allow for a table-top quantum simulation of quantum physics in curved background. Our mapping provides a computational tool to determine observables of the discrete system even for large lattices, where exact diagonalization fails. As an application and proof of principle we quantitatively reproduce the ground state energy, spectral gap, and correlation functions of the noninteracting lattice system by means of analytic formulas on the Poincar\'{e} disk, and show how conformal symmetry emerges for large lattices. This sets the stage for studying interactions and disorder on hyperbolic graphs in the future. Importantly, our analysis reveals that even relatively small discrete hyperbolic lattices emulate the continuous geometry of negatively curved space, and thus can be used to experimentally resolve fundamental open problems at the interface of interacting many-body systems, quantum field theory in curved space, and quantum gravity.
\end{abstract}

\maketitle

Non-Euclidean and hyperbolic geometry has inspired thinkers for millennia due to its intriguing properties and perplexing beauty \cite{Cannon}.
Besides its aesthetic appeal, the immense usefulness of hyperbolic geometry in physics due to the famous AdS/CFT correspondence \cite{maldacena1999large,witten1998anti} makes field theories and quantum physics in hyperbolic space one of the central themes of current theoretical research.
Furthermore, many recent developments in the context of holography and quantum information point towards a deep connection between geometry, entanglement, and renormalization \cite{PhysRevLett.96.181602,PhysRevLett.99.220405,PhysRevLett.101.110501,PhysRevD.86.065007,PhysRevLett.110.100402,milsted2018geometric}.
Other exciting applications of hyperbolic geometry emerge, for instance, in the field of fault-tolerant quantum computing \cite{Breuckmann_2016,Breuckmann_2017,Lavasani_2019}.
To elevate the study of quantum physics in hyperbolic space from pure theory to experimentally verifiable questions, it is crucial to create laboratory setups for exploring the underlying effects in a tunable manner.

Important progress towards the quantum simulation of curved space has been made in nonlinear optical media  \cite{PhysRevLett.84.822,PhysRevA.78.043821,genov2009mimicking,PhysRevLett.105.067402,Chen10,bekenstein2015optical,bekenstein2017control}, ultracold quantum gases \cite{PhysRevLett.85.4643,PhysRevLett.91.240407,PhysRevA.70.063615,carusotto2008numerical,PhysRevD.85.044046,Boada_2011,PhysRevLett.118.130404,Kosior_2018}, and other platforms \cite{PhysRevLett.95.031301,PhysRevLett.104.250403,PhysRevLett.106.021302,Sheng_2013,PhysRevD.95.125003}, which allowed, for instance, for observation of event horizons \cite{PhysRevLett.105.240401,Philbin_2008} and Hawking radiation \cite{Steinhauer_2016,Hu_2019}. In these experiments, curvature is often emulated in Euclidean geometries through nonlinear field propagation. A complementary path was followed in recent cutting-edge experiments in circuit quantum electrodynamics (QED) \cite{houck2012chip,PhysRevA.86.023837,schmidt2013circuit,PhysRevX.7.011016,PhysRevX.6.041043}, where hyperbolic geometry was emulated directly through photon dynamics confined to a hyperbolic lattice made from  superconducting resonators \cite{kollar2019hyperbolic,kollr2019linegraph}. The setup is highly tunable and can be used to achieve photon interactions, coupling to spin degrees of freedom, or the effects of disorder \cite{PhysRevB.82.100507,PhysRevX.4.031043,PhysRevX.9.011021}. Hyperbolic lattices have been investigated in the context of classical \cite{Rietman_1992,Shima_2006,PhysRevE.77.022104,Krcmar_2008,PhysRevE.79.011124,Gu_2012} and quantum spin systems \cite{Daniska_2016}, and complex networks \cite{PhysRevE.82.036106}.

In this work, we show that quantum many-body Hamiltonians relevant for circuit QED on hyperbolic lattices can be approximated by a continuum theory on the Poincar\'{e} disk. This provides a computational tool to access observables even for otherwise intractable very large lattices, and shows that the discrete setup constitutes a quantum simulation of continuous hyperbolic space. We provide a dictionary between discrete and continuous geometry. To show the strength of our mapping, we quantitatively reproduce the ground state energy, spectral gap, and correlation functions for the noninteracting theory by analytic continuum formulas. We reveal how conformal symmetry emerges on hyperbolic lattices, implying significant computational simplifications in applications.

\begin{figure}[t]
\centering
\includegraphics[width=8.5cm]{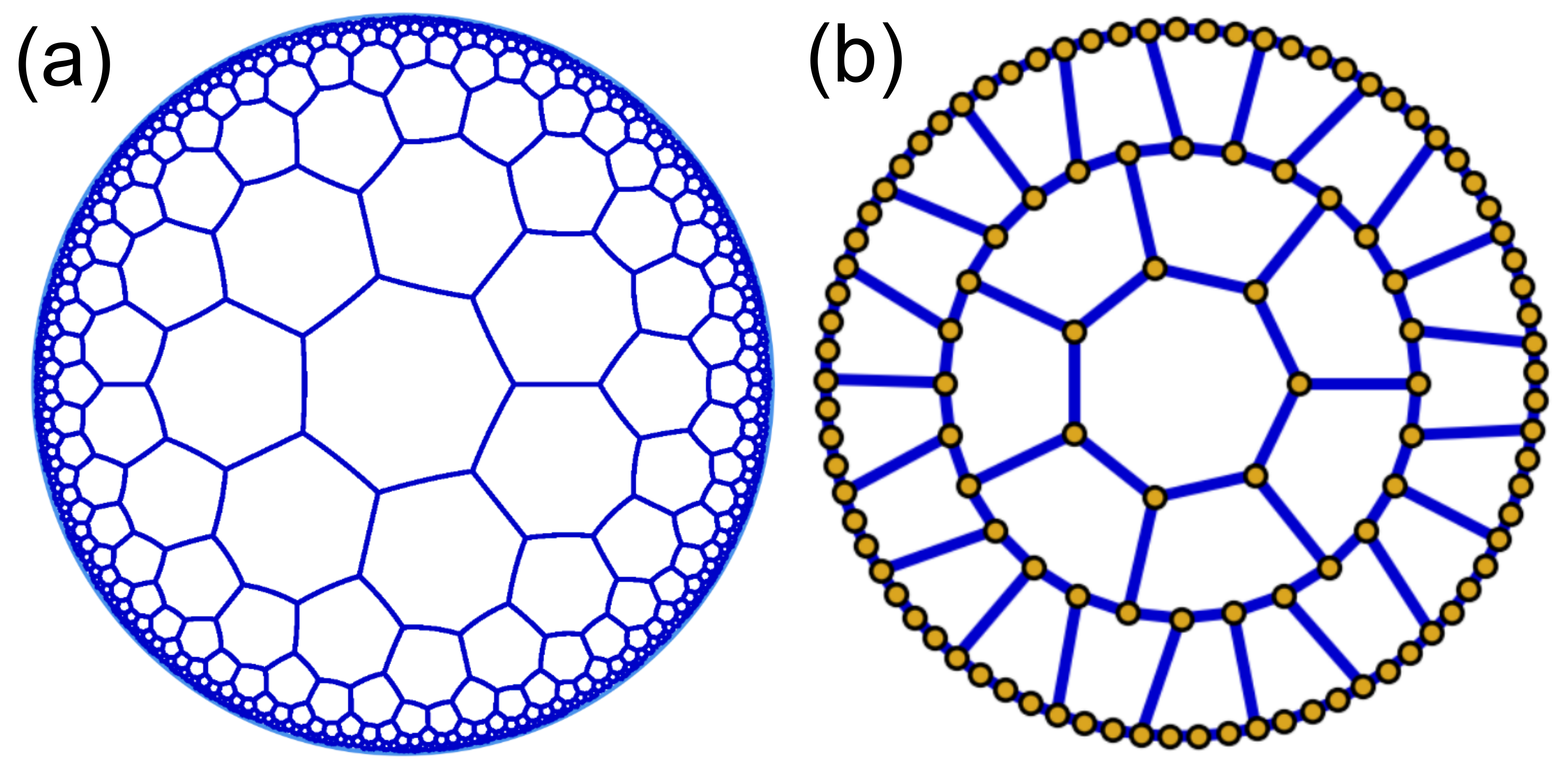}
\caption{(a) We consider the regular tessellation of the hyperbolic plane with heptagons, embedded into the Poincar\'{e} disk with a hyperbolic metric. All neighboring lattice sites have equal hyperbolic distance and the unit disk boundary is infinitely far away from each point in the interior. (b) Finite graphs preserving sevenfold rotation invariance can be constructed by considering subsets that are topologically equivalent to $\ell=1,2,3\dots$ concentric rings, shown here for $\ell=3$.}
\label{Fig1}
\end{figure}

\renewcommand{\arraystretch}{1.6}
\begin{table*}[t]
\begin{tabular}{|c||c|c|c|c|c|c|c|c|c|c|}
\hline number of rings $\ell$ & 1 & 2 & 3 & 4 & 5 & 6 & 7 & 8 & 9 & 10\\
 \hline\hline number of graph sites $N$ & 7 & 35 & 112 & 315 & 847 & 2240 & 5887 & 15435 & 40432 & 105875 \\
 \hline effective disk radius $L$ & 0.447 & 0.745 & 0.894 & 0.958 & 0.984 & 0.994 & 0.998 & 0.9990 & 0.9997 & 0.9999 \\
\hline\hline ground state energy (graph) $E_0$  & $-2$ & $ -2.636$ & $-2.787$ & $-2.847$ & $-2.877$ & $-2.894$ & $-2.905$ & $-2.91$   & $-2.92$ & $-2.92$ \\
 \hline ground state energy (continuum) $E_0^{(\rm cont)}$  & $-1.500$ & $-2.570$ & $-2.770$ & $-2.842$ & $-2.876$ & $-2.895$ & $-2.906$ & $-2.914$ & $-2.920$ & $-2.924$ \\
\hline
\end{tabular}
\caption{The total number of sites $N = 7[(\frac{3+\sqrt{5}}{2})^\ell+(\frac{3-\sqrt{5}}{2})^\ell-2]$ grows exponentially as a function of the number of rings $\ell$. Each finite graph is mapped onto a continuous disk of radius $L=\sqrt{N/(N+28)}$. The ground state energy $E_0$ of the hopping Hamiltonian (\ref{eq1}), defined as the lowest eigenvalue of the matrix $H=-A$, can be estimated from the lowest eigenvalue of $\Delta_g$ on the finite disk of radius $L<1$, which gives $E_0^{(\rm cont)}$. Both values agree excellently for sufficiently large $\ell$, see also Fig. \ref{Fig2}. For $\ell\geq 8$ we have to resort to less precise sparse matrix methods to estimate $E_0$.}
\label{Tab1}
\end{table*}
\renewcommand{\arraystretch}{1}

The photon dynamics of the circuit QED experiments of Ref. \cite{kollar2019hyperbolic} can be modeled as nearest-neighbor hopping on a hyperbolic graph $G$, see Fig. \ref{Fig1}. We label the graph in Schl\"{a}fli notation by $\{p,q\}$, which corresponds to a tessellation of the plane by regular $p$-gons with coordination number $q$. Whereas the three Euclidean lattices (triangular lattice $\{3,6\}$, square lattice $\{4,4\}$, honeycomb lattice $\{6,3\}$) satisfy $(p-2)(q-2)=4$, one can show that a hyperbolic tessellation is obtained for $(p-2)(q-2)>4$. In this work, we focus on the heptagonal hyperbolic lattice $\{p,3\}$ with $p=7$ for concreteness, but point out how to generalize the results to $p\geq 7$. The actual circuit QED experiments realize the line graph of $G$ \cite{kollr2019linegraph}, but their continuum approximation is analogous.

We consider the nearest-neighbor hopping Hamiltonian 
\begin{align}
 \label{eq1} \hat{\mathcal{H}}_0 = -t \sum_{i,j\in G} \hat{a}^\dagger_i A_{ij} \hat{a}_j,
\end{align}
with $\hat{a}_i^\dagger$ the photon creation operator on site $i$ and $t>0$. The entry $A_{ij}$ of the adjacency matrix $A$ is 1 if sites $i$ and $j$ are connected by an edge, and zero otherwise. We construct finite hyperbolic graphs that preserve sevenfold rotation invariance from $\ell=1,2,3,\dots$ successive quasi-concentric rings, where $\ell=1$ corresponds to a single heptagon, see Fig. \ref{Fig1}. The total number of sites grows exponentially as $N \sim 7 \vphi^{2\ell}$ for large $\ell$, with $\vphi=(1+\sqrt{5})/2$ the golden ratio \cite{PhysRevE.79.011124,Gu_2012}, see Table \ref{Tab1}. Sites in the interior of $G$ have coordination number 3, and sites on the boundary have either 2 or 3. The average coordination number for large $\ell$ is $3-1/\sqrt{5}= 2.553$ and there is always a significant fraction of boundary sites.

For the continuum description, we embed the hyperbolic lattice into the Poincar\'{e} disk $\mathbb{D}=\{z \in \mathbb{C}, |z|<1\}$ with hyperbolic metric
\begin{align}
 \label{eq3} \mbox{d}s^2 = \frac{\mbox{d}x^2+\mbox{d}y^2}{(1-|z|^2)^2}.
\end{align}
(We write $z=x+\rmi y = r e^{\rmi \phi}$.) Let us briefly recall some properties of this space of constant negative curvature \cite{Cannon}:
The hyperbolic distance between two points $z,z'\in\mathbb{D}$ is 
\begin{align}
\label{eq4} d(z,z') = \frac{1}{2} \text{arcosh}\Bigl(1+ \frac{2|z-z'|^2}{(1-|z|^2)(1-|z'|^2)}\Bigr),
\end{align}
which reduces to $|z-z'|$ for $|z|,|z'|\ll 1$. The boundary of $\mathbb{D}$ is infinitely far from every point in the interior and the area of a disk of radius $L<1$ is $\pi L^2/(1-L^2)$. The isometries (distance preserving maps) of $\mathbb{D}$ are given by conformal automorphisms
\begin{align}
 \label{eq5} z \mapsto w(z) = e^{\rmi \eta} \frac{a-z}{1-z \bar{a}}
\end{align}
with $\eta\in [0,2\pi)$ and $a\in \mathbb{D}$. These transformations exchange $a$ with the origin, and so each point in $\mathbb{D}$ is equivalent. The group of mappings (\ref{eq5}) is isomorphic to the group $\text{PSL}(2,\mathbb{R})$ of M\"{o}bius transformations on the upper half-plane. The embedding assigns a coordinate $z_i\in \mathbb{D}$ to each site $i\in G$ so that neighboring sites are separated by a hyperbolic distance $d_0=0.283128$ \cite{SMdata}, see Appendix \ref{AppGraph}. Importantly, the value of $d_0$ is determined by the lattice geometry and cannot be varied.

On the Euclidean square lattice, nearest-neighbor hopping Hamiltonians of type (\ref{eq1}) are related to the Laplacian through a finite difference approximation, facilitating powerful techniques such as the continuum theory of solids or lattice gauge theory in high-energy physics. The natural generalization of the Laplacian to curved manifolds is the Laplace--Beltrami operator
\begin{align}
 \label{eq7b} \Delta_g = \frac{1}{\sqrt{\mbox{det}g}} \partial_i\Bigl( \sqrt{\mbox{det}g}(g^{-1})_{ij}\partial_j\Bigr),
\end{align} 
where the metric tensor $g_{ij}=(1-r^2)^{-2}\delta_{ij}$ is related to the line element in Eq. (\ref{eq3}) by $\mbox{d}s^2=g_{ij}\mbox{d}x_i \mbox{d}x_j$. This operator is self-adjoint with respect to the canonical scalar product $\langle f_1,f_2\rangle = \int \mbox{d}V_g f_1^*f_2$, with $\mbox{d}V_g = \mbox{d}^2x \sqrt{\mbox{det}g}$ the invariant volume element. In our case, we find the \emph{hyperbolic Laplacian} to be
\begin{align}
 \label{eq7} \Delta_g = (1-|z|^2)^2 (\partial_x^2+\partial_y^2).
\end{align}
This operator is invariant under conformal automorphisms \cite{Cannon,sarnak2003spectra,marklof2012ii}.

\begin{figure}[t]
\centering
\includegraphics[width=8.5cm]{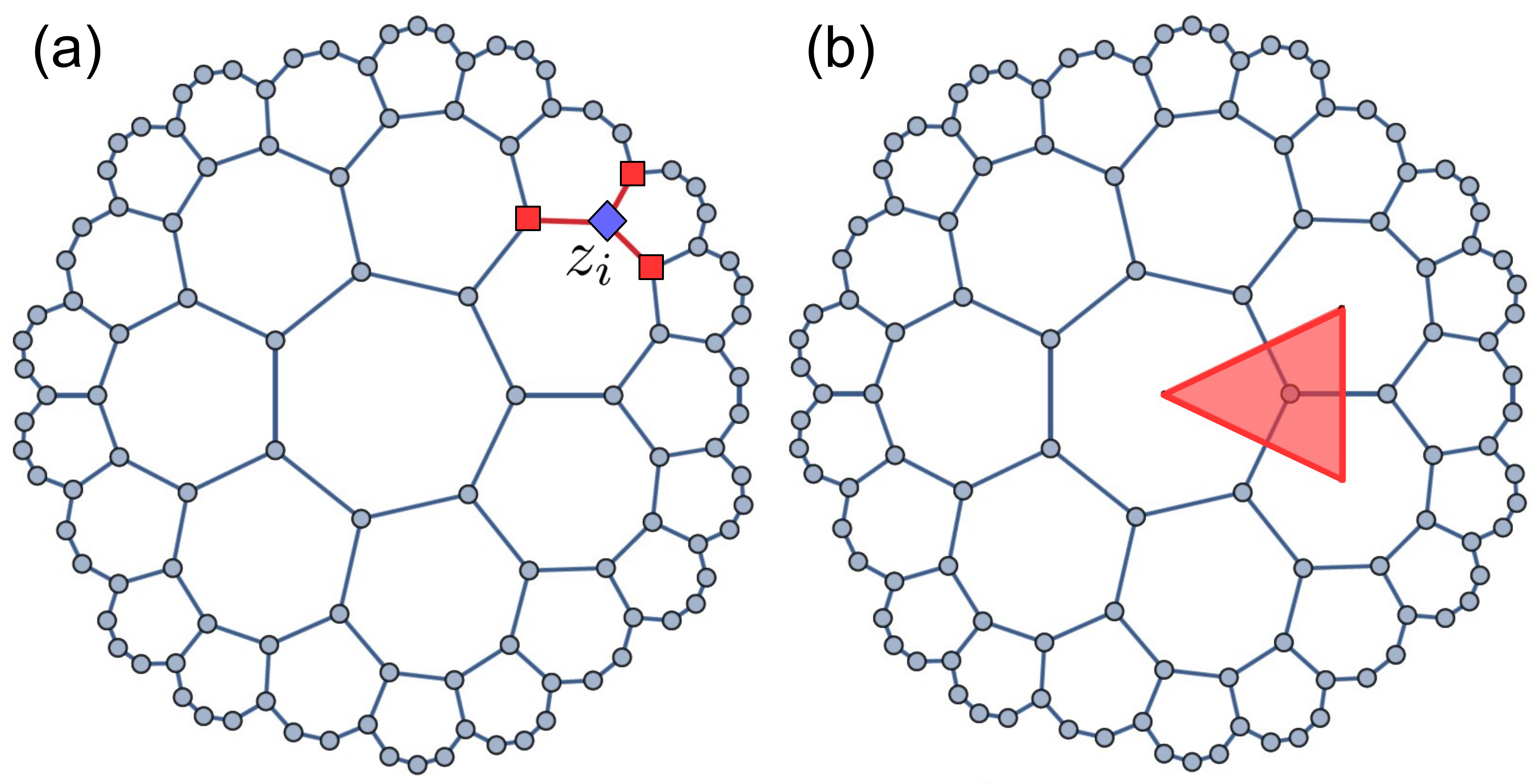}
\caption{(a) The adjacency matrix can be approximated by the hyperbolic Laplacian in the continuum limit through Eq. (\ref{eq10}). To derive this property, we choose an arbitrary site $z_i$ with coordination number 3 (blue diamond). When applying the automorphism $z \mapsto w(z) = \frac{z_i-z}{1-z\bar{z}_i}$, which maps $z_i$ to the origin, the three neighbors of $z_i$ (red squares) are mapped to an equilateral triangle. This implies Eq. (\ref{eq9}), which can be expanded in powers of $h$ to yield the desired relation. (b) Sums over lattice sites are replaced by integrals over hyperbolic space according to Eq. (\ref{eq11}). This is achieved by assigning to each site $z_i$ an effective hyperbolic triangle with interior angles $2\pi/7$ and area $\mathcal{A}_\bigtriangleup=\pi/28$.}
\label{FigNew}
\end{figure}

To understand why $\Delta_g$ appears here, note that every function $f:\mathbb{D} \to \mathbb{C}$ induces a function on the graph via $i \mapsto f(z_i)$. Consider then a lattice site $z_i$ with coordination number 3 and a sufficiently smooth function $f(z)$, see Fig. \ref{FigNew}. We have $A_{ij} f(z_j) = f(z_{i+e_1}) + f(z_{i+e_2}) + f(z_{i+e_3})$, where the right-hand side represents the sum over the neighbors of $z_i$ and we implicitly sum over repeated indices. To manipulate this expression, apply an automorphism $z\mapsto w(z)$ from Eq. (\ref{eq5}) with $\eta=0$ and $a=z_i$. This exchanges $z_i$ with the origin. Furthermore, the three neighbors of $z_i$ are mapped to form an equilateral triangle with coordinates $w_1 = h e^{\rmi \chi_i}$, $w_2= h e^{\rmi \chi_i}e^{\rmi 2\pi /3}$, $w_3=h e^{\rmi \chi_i}e^{\rmi 4\pi /3}$, and $h=\tanh(d_0)=0.275798$. The phase $\chi_i$ depends on the coordinate $z_i$ in a nontrivial manner, see Fig. \ref{FigScos3delta} in the Appendix. Applying the inverse automorphism we arrive at the identity
\begin{align}
 \label{eq9} A_{ij}f(z_j) = f\Bigl(\frac{z_i-w_1}{1-w_1\bar{z}_i}\Bigr) + f\Bigl(\frac{z_i-w_2}{1-w_2\bar{z}_i}\Bigr) + f\Bigl(\frac{z_i-w_3}{1-w_3\bar{z}_i}\Bigr).
\end{align}
This equation can be expanded in powers of $h$, with the linear term vanishing due to $w_1+w_2+w_3=0$, and the quadratic term being universal and independent of $\chi_i$:
\begin{align}
 \label{eq10} A_{ij} f(z_j) = 3f(z_i)  + \frac{3}{4}h^2 \Delta_g f(z_i) + \mathcal{O}(h^3).
\end{align}
This relation between the adjacency matrix and the hyperbolic Laplacian remains true for $p$-gons with $p>7$, albeit with a different value of $h$. We emphasize again that $h=\tanh(d_0)$ is fixed by hyperbolic geometry and cannot be tuned. However, the right-hand side of Eq. (\ref{eq9}) can be evaluated for every $h$ and so permits a formal finite size scaling limit $h\to 0$. In Appendix \ref{AppAdj} we compute the $\mathcal{O}(h^3)$-correction to Eq. (\ref{eq10}) and discuss the role of boundary sites. In Appendix \ref{AppFS}, we show that it diminishes as the lattice parameter decreases.

The next step towards a continuum theory for hyperbolic lattices is a formula to approximate sums over lattice sites by integrals over the Poincar\'{e} disk. For suitable functions $f:\mathbb{D}\to \mathbb{C}$, an argument from finite-element triangulations \cite{BookFEM,PhysRevD.98.014502,PhysRevD.95.114510,brower2019lattice} implies
\begin{align}
 \label{eq11} \sum_{i\in G} f(z_i) \approx \frac{1}{\mathcal{A}_\bigtriangleup} \int_{|z|\leq L} \frac{\mbox{d}^2z}{(1-|z|^2)^2}f(z).
\end{align}
Here, $\mbox{d}V_g=\mbox{d}^2z/(1-|z|^2)^2$ is the invariant hyperbolic volume element and $\mathcal{A}_\bigtriangleup=(\pi-\gamma_1-\gamma_2-\gamma_3)/4=\pi/28$ is the area of the hyperbolic triangle of the dual lattice with interior angles $\gamma_1=\gamma_2=\gamma_3=2\pi/7$. Importantly, this implies that a finite graph with $\ell$ rings and $N(\ell)$ sites corresponds to a finite continuous disk with effective radius
\begin{align}
 \label{eq12} L = \sqrt{\frac{N}{N+28}},
\end{align}
the value of which is determined such that the right-hand side of Eq. (\ref{eq11}) yields $N$ when inserting $f=1$. We display the first ten effective radii in Table \ref{Tab1}. In Appendix \ref{AppSum}, we present an alternative derivation of Eq. (\ref{eq11}), which does not utilize the dual lattice. For tessellations with $p$-gons with $p\geq 7$, we replace $28 \to \pi/\mathcal{A}_\bigtriangleup =4p/(p-6)$.

This dictionary how to approximate $A_{ij}$ and $\sum_{i\in G}$ on the hyperbolic lattice by their continuum counterparts comprises the first main result of this work. As an example consider the Bose--Hubbard model Hamiltonian on the hyperbolic lattice
\begin{align}
 \label{eq13} \hat{\mathcal{H}}=  \sum_{i\in G} \Bigl[-t\sum_{j\in G} \hat{a}^\dagger_iA_{ij}\hat{a}_j  -\mu \hat{a}^\dagger_i \hat{a}_i+ U (\hat{a}^\dagger_i\hat{a}_i)^2\Bigr],
\end{align}
with chemical potential $\mu$ and on-site interaction $U$. An exciting quantum simulation challenge would be to understand the phase diagram and universality class of the superfluid-to-Mott insulator transition in this model. The associated superfluid in hyperbolic space is then captured by the continuum Hamiltonian
\begin{align}
 \label{eq14} \hat{\mathcal{H}}' = \int_{|z|\leq L}\frac{\mbox{d}^2z}{(1-|z|^2)^2}\Bigl[ \hat{\alpha}^\dagger_z(-t'\Delta_g -\mu')\hat{\alpha}_z+ U' (\hat{\alpha}^\dagger_z\hat{\alpha}_z)^2\Bigr]
\end{align}
with adjusted parameters $t'=\frac{3}{4}h^2t$, $\mu'=\mu+3t$, $U'=\frac{\pi}{28}U$. The field operators $\hat{\alpha}_{z}=\hat{\alpha}(z)$ satisfy curved-space commutation relations, $[\hat{\alpha}(z),\hat{\alpha}^\dagger(z')]=(1-|z|^2)^2\delta^{(2)}(z-z')$, and we have $\hat{\alpha}(z_i)=\sqrt{28/\pi}\hat{a}_i$. More generally, every many-body system with Lagrangian $\mathcal{L}(a_i)$ where the kinetic term results from nearest-neighbor hopping, especially including multiple species of both bosons and fermions and strong or long-ranged interactions, can be simulated in hyperbolic space by putting it onto a hyperbolic lattice. The resulting continuum theory is described by the action $S = \int\mbox{d}V_g\ \mathcal{L}(\alpha_z)$. In this work, we deliberately ignore boundary effects, although holographic models and simulation of bulk-boundary dualities \cite{Son_2008,Balasubramanian_2008,axenides2014modular,axenides2018quantum,boyle2018conformal} are a fascinating application that we leave for future work.

\begin{figure}[t]
\centering
\includegraphics[width=8.5cm]{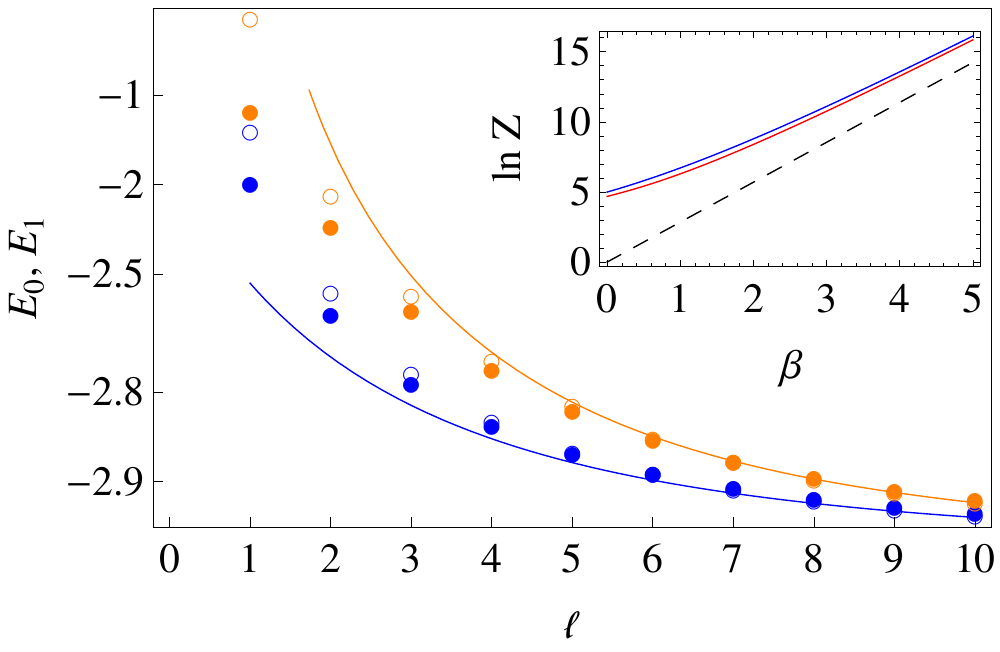}
\caption{Ground state energy $E_0$ (blue, lower data) and first excited state energy $E_1$ (orange, upper data) for the graph (filled circles) and corresponding continuous disk (empty circles). The solid lines are the asymptotic continuum formulas from Eq. (\ref{eq19}). \emph{Inset:} Partition function $Z$ summed over negative energies for the graph with $\ell=4$ (blue) and associated disk (red) as a function of inverse temperature $\beta$. The dashed line is the low-temperature asymptote $\ln Z \sim -\beta E_0$.}
\label{Fig2}
\end{figure}

The spectral theory of the hyperbolic Laplacian on the Poincar\'{e} disk is well-understood \cite{sarnak2003spectra,marklof2012ii}. We summarize the main aspects here and give some additional details in Appendix \ref{AppEig}. The eigenvalues of $-\Delta_g$ are $\vare = 1+k^2$ with eigenfunctions
\begin{align}
   \label{eq16a} \psi_K(z) = \Bigl(\frac{1-|z|^2}{|1-z e^{-\rmi \beta}|^2}\Bigr)^{\frac{1}{2}(1+\rmi k)},
\end{align}
where $K=k e^{\rmi \beta}$ is a two-dimensional momentum vector. For $z\to 0$ and $k\to \infty$ (corresponding to infinite disk radius and $\Delta_g\to \Delta$) we have $\vare \to k^2$ and $\psi_K(z) \to e^{\frac{\rmi}{2}(K\bar{z}+z\bar{K})} = e^{\rmi \textbf{k}\cdot \textbf{x}}$, which is the local Euclidean plane wave limit. In radial coordinates, we write
\begin{align}
   \label{eq16b} \psi_K(z) = \sum_{m=-\infty}^\infty \rmi^m g_{km}(r) e^{\rmi m(\phi-\beta)},
\end{align}
generalizing the partial wave decomposition of plane waves. We have
\begin{align}
  \label{eq16} g_{km}(r) \propto P_{\frac{1}{2}(-1+\rmi k)}^m\Bigl(\frac{1+r^2}{1-r^2}\Bigr),
\end{align}
where $P_\nu^m$ is the Legendre function of the first kind. Restricting space to a finite disk of radius $L<1$ and imposing Dirichlet boundary conditions, $\psi_{km}(L)=0$, we obtain a discrete energy spectrum $\vare_{n}=1+k_{n}^2$ with $k_n>0$, analogous to a particle in a spherical well potential

As a first application of the continuum theory, we estimate ground state energy and spectral gap of the Hamiltonian in Eq. (\ref{eq1}). We set $t=1$. Since $\hat{\mathcal{H}}_0$ is quadratic this reduces to determining the lowest two eigenvalues of the matrix $H=-A$, which we label $E_0$ and $E_1 = E_0 +\delta E$ with spectral gap $\delta E>0$. Note that the spectrum of $H$ is contained in the real interval $(-3,3)$. For infinite lattices, the ground state energy is known from mathematical graph theory to lie in the interval \cite{Paschke,Higuchi,kollr2019linegraph} $\lim_{\ell \to \infty}E_0 \in [-2.966,-2.862]$.
To estimate the lowest eigenvalues from the continuum limit, we approximate $H$ by the differential operator 
\begin{align}
H^{\rm (cont)} = -3-\frac{3}{4}h^2 \Delta_g
\end{align}
with Dirichlet boundary conditions at radius $L$. Its eigenvalues are the discrete set $E_{n}^{(\rm cont)}=-3+\frac{3}{4}h^2(1+k_{n}^2)$. As $\ell \to \infty$, the lowest possible $k_n\to 0$, which yields the ground state energy $E_\infty = -3 + \frac{3}{4}h^2 = -2.94295$, consistent with the graph bound. For finite $\ell\geq 1$, the first two eigenvalues of $H^{(\rm cont)}$ are readily computed from $\psi_{km}(L)=0$ for $m=0,1$, respectively. They agree remarkably well with the graph data for $\ell\gtrsim 4$, see Table \ref{Tab1} and Fig. \ref{Fig2}. For $\ell\to \infty$ we have
\begin{align}
 \label{eq19} E_{0,1}^{(\rm cont)} \sim E_\infty + \frac{3\pi^2h^2}{4}\frac{1}{(\ln\vphi\cdot \ell+c_{0,1})^2},
\end{align}
with $\vphi=(1+\sqrt{5})/2$, $c_0=\ln 2$, and $c_1=\ln2-1$. Equation (\ref{eq19}) is derived in Appendix \ref{AppSpec}.

As we go to higher energies, the spectra of $H$ and $H^{(\rm cont)}$ start to deviate. Still, the graph and continuum partition functions $Z=\sum_n\Theta(-E_n) e^{-\beta E_n}$ and $Z'= \sum_{n}\Theta(-E_n^{(\rm cont)})e^{-\beta E_{n}^{(cont)}}$ with inverse temperature $\beta$ agree well, see the inset of Fig. \ref{Fig2}. (We limit the sums to negative energies to roughly cut off high energy contributions clearly outside the continuum approximation.) The ability to \emph{quantitatively} reproduce the low-energy graph spectrum and predict the behavior for large graphs by means of the continuum approximation constitutes the second main result of this work.

\begin{figure}[t]
\centering
\includegraphics[width=8.5cm]{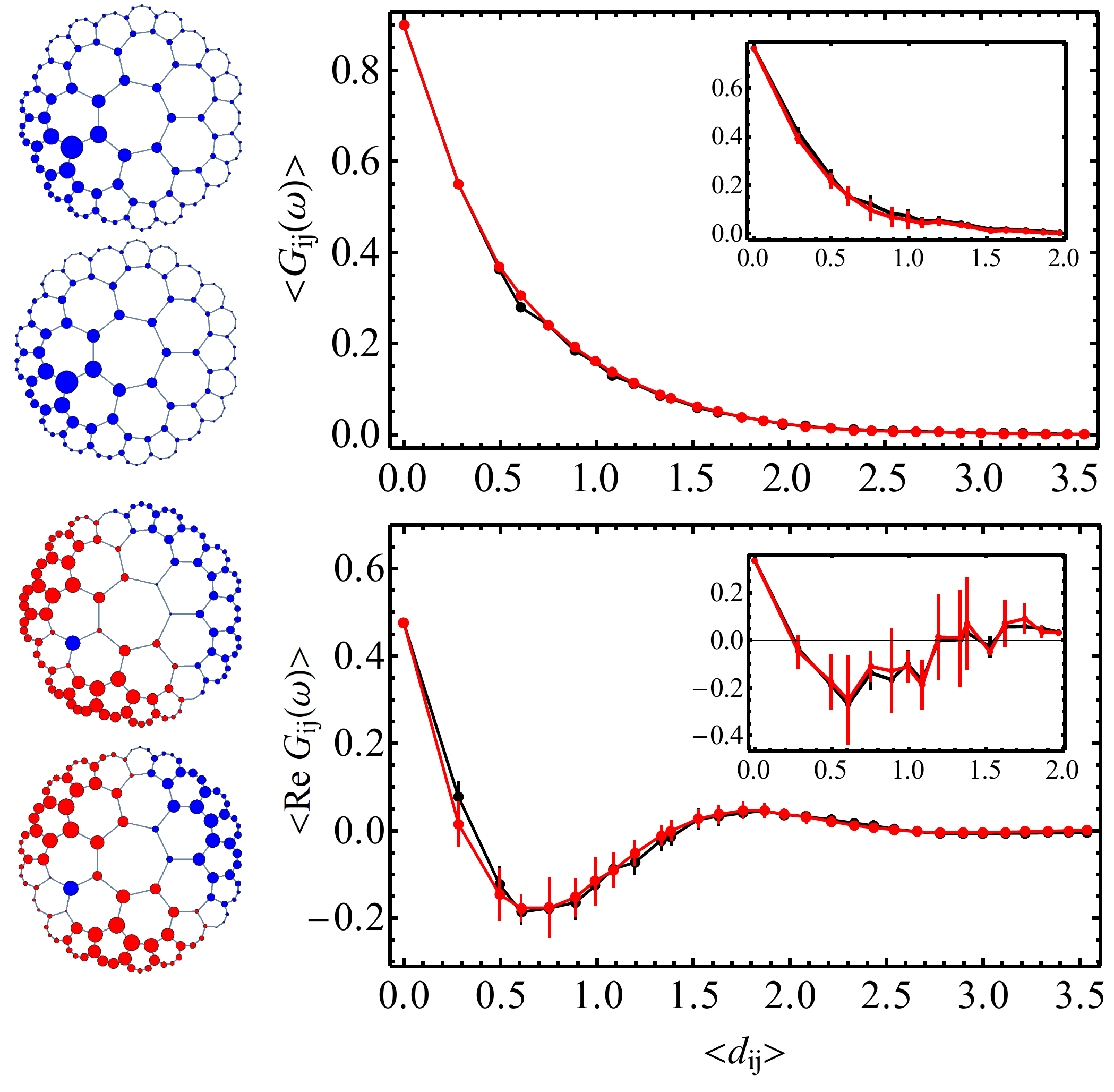}
\caption{Quantitative match between graph Green function $G_{ij}$ and continuum Green function $G(z_i,z_j)$. We fix site $z_i$ to be on the 2nd ring, and plot the correlations as a function $F_j$ of site $z_j$. \emph{Upper panel.} Results for $\omega=-2.95$ just below $E_0$. \emph{(Left).} The two plots are $F_j=G_{ij}$ and $F_j=G(z_i,z_j)$. The size of dots is proportional to $|F_j|^{1/2}$, and blue/red corresponds to positive/negative sign of $F_j$. \emph{(Right).} Mean correlation function vs. hyperbolic distance $d_{ij}=d(z_i,z_j)$, where the red (black) data is the graph (continuum) function. To obtain the curves, we make a list of pairs $(F_j,d_{ij})$ and compute the average $F_j$ as a function of distance, with the error bar being the standard deviation. The quantitative agreement between graph and continuum is remarkable. Emergent conformal symmetry is reflected by the data points collapsing onto a single curve $G_{ij}=f(d_{ij})$ with some function $f$ for large $\ell$. The main plots are for $\ell=6$, the insets for $\ell=3$. \emph{Lower panel.} The same setting for $\omega=-2.5+ 0.1\rmi$ with $\text{Re}(\omega) >E_0$. We plot the real part of the correlation function.}
\label{Fig3}
\end{figure}

Our second application of the continuum theory is the computation of correlation functions on the graph from the continuum Green function. We denote the Green function of $H=-A$ by
\begin{align}
 \label{eq20} G_{ij}(\omega) = \Bigl(\frac{1}{H-\omega}\Bigr)_{ij} = \sum_{n=1}^N \frac{\psi_n(i)\psi_n^*(j)}{E_n-\omega}.
\end{align}
Here $\psi_n$ and $E_n$ are the eigenvectors and eigenenergies of $H$, $H\psi_n = E_n \psi_n$, and $\omega\in \mathbb{C}$ is a complex frequency. $G_{ij}(\omega)$ constitutes the measurable two-point correlation function $\langle \hat{a}_i(\omega)\hat{a}_j^\dagger(\omega)\rangle_0$ for the free theory averaged with respect to $\hat{\mathcal{H}}_0$, and can be written as an auxiliary field Gaussian path integral on the graph, see Appendix \ref{AppPath}. Approximating the latter by the continuum expressions we obtain
\begin{align}
 \label{eq21} G_{ij}(\omega) \approx \frac{\pi}{21 h^2}\ G\Bigl(z_i,z_j, \frac{4(\omega+3)}{3h^2},L\Bigr).
\end{align}
Here $G(z,z',\lambda,L)$ is the Green function of the hyperbolic Helmholtz operator, i.e. it satisfies
\begin{align}
(\lambda+\Delta_g)G(z,z',\lambda,L) = -(1-|z|^2)^2 \delta^{(2)}(z-z')
\end{align}
and Dirichlet boundary conditions $G(z,z',\lambda,L)=0$ for $|z|=L$ or $|z'|=L$. Again, the disk radius $L$ is matched to $\ell$ through Eq. (\ref{eq12}). The accuracy of the approximation in Eq. (\ref{eq21}) is remarkably good as shown in Fig. \ref{Fig3}.

The continuum Green function $G(z,z',\lambda,L)$ is uniquely specified by the Dirichlet boundary condition \cite{AbrStegun,ZinnJustinBook,marklof2012ii,PhysRevA.94.011602}. The full but lengthy analytic expression is derived in Appendix \ref{AppGreen} and summarized in Appendix \ref{AppGreenSum}. As $L\to 1$, the Green function is solely a function of the hyperbolic distance $d(z,z')$ due to automorphism invariance. For $\lambda=0$ we have
\begin{align}
 \label{eq23} G(z,z',0,L) = -\frac{1}{4\pi}\ln\Bigl|\frac{L(z-z')}{L^2-z\bar{z}'}\Bigr|^2,
\end{align}
which, indeed, is a function of  $\tanh d(z,z')=|\frac{z-z'}{1-z\bar{z}'}|$ for $L=1$. In turn, this implies that also the graph correlation function $G_{ij}(\omega)$ is approximately a universal function of the hyperbolic distance $d_{ij}=d(z_i,z_j)$ for large $\ell$, as is shown in Fig. \ref{Fig3}.  The \emph{quantitative} matching of graph and continuum Green functions and the finding of emergent conformal symmetry on the hyperbolic lattice constitute our third main result.

The continuum approximation for hyperbolic lattices that we have put forward shows a path how quantum many-body systems in continuous hyperbolic space can be reliably simulated using discrete geometries. It also gives a computational framework to efficiently compute observables relevant for experiments simulating curved spaces. In particular, the continuum Green function can be used in diagrammatic techniques to accurately determine interaction effects even for relatively small system sizes. One platform of interest is, of course, circuit QED realizations, where the interplay of curvature and strong interactions between photons and qubit degrees of freedoms can be studied. However, hyperbolic lattices or their topological equivalents can, in principle, also be realized in other experimental platforms that allow for engineered hopping on graphs, including optical lattices for ultracold atoms \cite{Gross995} or electrical circuits \cite{Lee2018}, and thus can be utilized to simulate other many-body Hamiltonians in hyperbolic space (including fermions or artificial gauge fields). The strong coupling regime in experiment has the potential to uncover novel effects and can be used to benchmark our theoretical description of nonperturbative quantum physics in curved space.

The continuum formalism also naturally connects to statements of AdS/CFT correspondence and the intriguing interplay between boundary field theories and a gravitational bulk. In this context, we want to make the point that simulating bosons in hyperbolic space is related to simulating quantum gravity in two Euclidean dimensions. Although the Einstein--Hilbert action is purely topological in two dimensions, an important alternative theory for metric fluctuations is Liouville quantum gravity \cite{ginsparg1993lectures}, which also appears in the context of the SYK model \cite{Bagrets2016} and bosonic string theory \cite{BookString}. Using that every two-dimensional metric can be written as $g_{ij} = e^{\varphi}\delta_{ij}$ with a scalar field $\varphi$, a saddle-point expansion of the Liouville action yields a field theory for $\varphi$ in a hyperbolic background of constant negative curvature, and fluctuations of $\varphi$ in that background correspond to fluctuations of the metric $g_{ij}$. We conclude that hyperbolic lattices promise a bright future for the genuine simulation of quantum physics in curved space.

\emph{Acknowledgments.} We gratefully acknowledge collaboration and many inspiring discussions with R. Lundgren and A. Houck. We thank T. Jacobsen, V. Galitski, B. Swingle, and Y. Wang for insightful comments.  This work was supported by DOE BES award DE-SC0019449 (hyperbolic lattice generation and analytical results), by the United States Army Research Lab’s Center for Distributed Quantum Information (CDQI) at the University of Maryland (numerical implementation), and by the National Science Foundation Physics Frontier Center at the Joint Quantum Institute award PHYS-1430094 (applications and asymptotic limits). R.B. also acknowledges fellowship support from NSERC and FRQNT. R. B. acknowledges support from NSERC and FRQNT.

\begin{appendix}

\section{Embedding coordinates}\label{AppGraph}

In this section, we explain how the finite graph with $\ell$ rings is embedded into the Poincar\'{e} disk.

We assign a coordinate $z_i\in\mathbb{D}$ to each graph site $i\in G$ such that neighboring sites are at hyperbolic distance $d_0$, with $d_0$ to be determined. In order to label the coordinates $z_i\in \mathbb{D}$, we have in mind the topologically equivalent graph with $\ell$ concentric rings, See Fig. 1 in the main text. We enumerate the $N$ graph sites with an index $i$ in a counterclockwise manner by starting on the first ring, then the second ring, and so forth. In this way, the first $i=1,\dots,7$ sites are on ring $\ell=1$, sites $i=8,\dots,35$ are on ring $\ell=2$, etc. The number of sites on each ring are summarized in Table \ref{AppTab1}.

The construction of the tessellation starts with the central regular heptagon with $|z_1|=\dots|z_7|=r_0$. For a general hyperbolic lattice $\{p,q\}$ we have
\begin{align}
 \label{Eqr0} r_0 = \sqrt{\frac{\cos(\frac{\pi}{q}+\frac{\pi}{p})}{\cos(\frac{\pi}{p}-\frac{\pi}{q})}}
\end{align}
and so for our tessellation $\{7,3\}$ we find
\begin{align}
 r_0 =  0.300743.
\end{align}
In particular, the first two coordinates are $z_1=r_0$ and $z_2= r_0 e^{2\pi \rmi/7}$ so that
\begin{align}
 d_0 = d(z_1,z_2) = 0.283128
\end{align}
with $d(z,z')$ the hyperbolic distance in the Poincar\'{e} disk. Starting from the central heptagon, the hyperbolic lattice is generated by iteratively applying the two generators of the symmetry group of the tessellation to the existing sites, which are rotations by $2\pi/7$ through the center of a heptagon and rotations by $2\pi/3$ through a vertex. Alternatively, we can create polygons by iteratively inverting existing polygons on hyperbolic circles along the edges of the polygon.

A list of coordinates $\{z_i\}$ for the sites of the first $\ell=6$ rings is attached to this work as supplementary data in the file ``coordinates.dat'' \cite{SMdata}. The file contains 2240 lines, which correspond to the 2240 coordinates for the graph with six rings. The first (second) column of the data constitutes the real (imaginary) part of $z_i\in\mathbb{D}$. In order to extract the coordinates for a graph with $\ell=1,2,3,4,5,6$ rings, restrict to the first $7,35,112,315,847,2240$ lines of the data, respectively.

\begin{table}[t]
\begin{tabular}{|c|c|c|c|c|c|c|c|c|c|c|}
\hline $\ell$ rings & \ 1\ & 2 & 3 & 4 & 5 & 6 & 7 & 8 \\ 
\hline\hline\ $N_{\rm ring}(\ell)$ \ &7 & 28 & 77 & 203 & 532 & 1393 & 3647 & 9548 \\ 
 \hline $N(\ell)$ & 7 & 35 & 112 & 315 & 847 & 2240 & 5887 & 15435  \\ 
\hline 
\end{tabular} 
\caption{Number of sites on the $\ell$th ring, $N_{\rm ring}(\ell)$, and total number of sites for a graph with $\ell$ rings, $N(\ell)$, for the first eight rings.}
\label{AppTab1}
\end{table}

\section{Derivation of Eq. (7) and corrections}\label{AppAdj}

In this section we present a more detailed derivation of Eq. (7) for approximating the adjacency matrix by the hyperbolic Laplacian. We further discuss the validity of this relation for boundary sites with coordination number 2 and compute the next-to-leading order correction in the expansion in powers of $h$.

\emph{Derivation of Eq. (7).} Choose an arbitrary site $z_i$ of the hyperbolic lattice with coordination number 3. For a test function $f:\mathbb{D}\to\mathbb{C}$ such that $f(z_i)=f_i$ we then have
\begin{align}
 \label{adj1} A_{ij}f_j = f(z_{i+e_1}) +  f(z_{i+e_2}) +  f(z_{i+e_3}),
\end{align}
where $z_{i+e_a}$ with $a=1,2,3$ stands for the sites adjacent to $z_i$. The heptagonal lattice is such that all adjacent lattice sites have the same distance with respect to the hyperbolic metric. In particular, this property remains invariant under automorphisms of the Poincar\'{e} disk.  We apply the transformation $\mathbb{D}\to \mathbb{D}$,
\begin{align}
 \label{adj2} z \mapsto w(z) &= \frac{z_i-z}{1-z\bar{z}_i},\\
 \label{adj3} w \mapsto z(w) &=  \frac{z_i-w}{1-w\bar{z}_i},
\end{align}
which exchanges $z_i$ and the origin. We write  $f(z)=\tilde{f}(w(z))$ and have 
\begin{align}
 \label{adj4} A_{ij}f_j &=  \tilde{f}(w(z_{i+e_1})) + \tilde{f}(w(z_{i+e_2})) + \tilde{f}(w(z_{i+e_3})).
\end{align}
The adjacent sites in the rotated frame, however, have very simple coordinates: Modulo rotation, they correspond to three sites at hyperbolic distance $d_0$ from the origin, with mutual relative angle $2\pi/3$. The corresponding Euclidean distance $h$ in the disk is such that
\begin{align}
 \label{adj5} d(h,0)\stackrel{!}{=} d_0,
\end{align}
and so
\begin{align}
 \label{adj6} h = \tanh(d_0)=0.275798.
\end{align}
We write
\begin{align}
 \label{adj7} w_1 = w(z_{i+e_1}) &= he^{\rmi \chi_i},\\
 \label{adj8} w_2 =w(z_{i+e_2}) &= h e^{\rmi 2\pi/3}e^{\rmi \chi_i},\\
 \label{adj9} w_3 =w(z_{i+e_3}) &= h e^{\rmi 4\pi/3}e^{\rmi \chi_i},
\end{align}
where the angle $\chi_i$ is determined by the coordinate $z_i$, see below. We have
\begin{align}
 \label{adj10} w_1+w_2+w_3 = w_1^2+w_2^2+w_3^2=0.
\end{align}
Applying the inverse automorphism we can parametrize the sites adjacent to $z_i$ as
\begin{align}
 \label{adj11} z_{i+e_1} &= z(w_1) = \frac{z_i-w_1}{1-w_1\bar{z}_i},\\
 \label{adj12} z_{i+e_2} &= z(w_2) = \frac{z_i-w_2}{1-w_2\bar{z}_i},\\
 \label{adj13} z_{i+e_3} &= z(w_3) = \frac{z_i-w_3}{1-w_3\bar{z}_i}
\end{align}
and so
\begin{align}
 \label{adj14} A_{ij}f_j &= f\Bigl(\frac{z_i-w_1}{1-w_1\bar{z}_i}\Bigr) + f\Bigl(\frac{z_i-w_2}{1-w_2\bar{z}_i}\Bigr) + f\Bigl(\frac{z_i-w_3}{1-w_3\bar{z}_i}\Bigr).
\end{align}
The right-hand side is a complex number that depends on the parameter $h$ and can be approximated through Taylor's formulas by a polynomial in $h$. We write
\begin{align}
 \label{adj15} A_{ij} f_j = 3f(z_i) + Q_1 h + Q_2 h^2 +\mathcal{O}(h^3)
\end{align}
with
\begin{align}
 \nonumber Q_1 = \frac{\mbox{d}}{\mbox{d}h}\Bigl[{}&f\Bigl(\frac{z_i-w_1}{1-w_1\bar{z}_i}\Bigr)+ f\Bigl(\frac{z_i-w_2}{1-w_2\bar{z}_i}\Bigr)\\
 \label{adj16} &+ f\Bigl(\frac{z_i-w_3}{1-w_3\bar{z}_i}\Bigr)\Bigr]_{h=0},\\
 \nonumber Q_2 = \frac{1}{2}\frac{\mbox{d}^2}{\mbox{d}h^2}\Bigl[{}&f\Bigl(\frac{z_i-w_1}{1-w_1\bar{z}_i}\Bigr)+ f\Bigl(\frac{z_i-w_2}{1-w_2\bar{z}_i}\Bigr)\\
 \label{adj17} &+ f\Bigl(\frac{z_i-w_3}{1-w_3\bar{z}_i}\Bigr)\Bigr]_{h=0}.
\end{align}
We use a complex notation where we identify $f(z) \equiv f(z,\bar{z})$ and 
\begin{align}
\label{adj18}  \partial_z &= \frac{\partial}{\partial z} = \frac{1}{2}(\partial_x-\rmi \partial_y),\\
 \label{adj18b} \bar{\partial}_z &= \frac{\partial}{\partial \bar{z}} =\frac{1}{2}(\partial_x+\rmi \partial_y).
\end{align}
For $a=1,2,3$ (with $w_{a,h}:=\mbox{d} w_a/\mbox{d}h = w_a/h$) we arrive at
\begin{align}
 \nonumber &\frac{\mbox{d}}{\mbox{d}h} f\Bigl(\frac{z_i-w_a}{1-w_a\bar{z}_i}\Bigr)_{h=0}\\
 \label{adj22} &= -(1-|z_i|^2) \Bigl(w_{1,h}\partial_z+\bar{w}_{1,h}\bar{\partial}_z \Bigr)f(z_i),\\
 \nonumber &\frac{\mbox{d}^2}{\mbox{d}h^2} f\Bigl(\frac{z_i-w_a}{1-w_a\bar{z}_i}\Bigr)_{h=0} \\
 \nonumber &= (1-|z_i|^2)^2\Bigl(\partial_z^2f(z_i)w_{1,h}^2+\bar{\partial}_z^2f(z_i) \bar{w}_{1,h}^2\Bigr)\\
 \nonumber &-2\bar{z}_i(1-|z_i|^2)w_{1,h}^2\partial_zf(z_i)-2z_i(1-|z_i|^2)\bar{w}_{1,h}^2\bar{\partial}_zf(z_i)\\
 \label{adj23} & +2(1-|z_i|^2)^2 |w_{1,h}|^2\partial_z\bar{\partial}_zf(z_i).
\end{align}
Summing over the index $a$ and using $\sum_{a=1}^3 w_a = \sum_{a=1}^2w_a^2=0$ and $|w_{a,h}|=1$ this implies
\begin{align}
 \label{adj24} Q_1=0 
\end{align}
and 
\begin{align}
 \label{adj25} Q_2 &= 3(1-|z_i|^2)^2\partial_z \bar{\partial}_z f(z_i).
\end{align}
Note that $\Delta=4 \partial_z\bar{\partial}_z$. Thus we have shown that
\begin{align}
 \label{adj26} A_{ij}f_j = 3 f(z_i) + \frac{3}{4}h^2\Delta_g f(z_i) +\mathcal{O}(h^3)
\end{align}
for a site with coordination number 3.

\emph{Coordination number 2.} Any site $z_i$ with coordination number 2 necessarily lies on the boundary of the graph, which we assume to be built from $\ell$ rings. Denote the two sites adjacent to $z_i$ in $G$ by $z_{i+e_1}$ and $z_{i+e_2}$. The coordinate of the third neighboring site $z_{i+e_3}$ lies on the $(\ell+1)$th ring outside $G$, but is otherwise uniquely specified by the heptagonal tessellation of the hyperbolic plane. We then have
\begin{align}
 \nonumber A_{ij}f(z_i) &= f(z_{i+e_1}) + f(z_{i+e_2})\\
 \nonumber &= \underbrace{\Bigl[-f(z_{i+e_3}) + f(z_{i+e_3})\Bigr]}_{0} + f(z_{i+e_1}) + f(z_{i+e_2})\\
 \label{adj27} &= -f(z_{i+e_3}) + 3 f(z_i) +\frac{3}{4}h^2 \Delta_g f(z_i) +\mathcal{O}(h^3).
\end{align}
We could now expand the first term to linear order in $h$, giving $f(z_{i+e_3})=f(z_i)+ h\cdot\delta_n f(z_i)$, with $\delta_n f(z_i)$ a directional derivative along the line from $z_i$ to $z_{i+e_3}$. On the other hand, for our purposes we need Eq. (\ref{adj26}) only for the case that $f_i = \hat{a}_i = \hat{\alpha}(z_i)$ is an annihilation operator. For any many-body state $|\Psi\rangle$ describing the system on graph $G$ we have $\hat{\alpha}_{i+e_3}|\Psi\rangle =0$, since $z_{i+e_3}\notin G$. Consequently
\begin{align}
 A_{ij}\hat{a}_j|\Psi\rangle = \Bigl( 3 \hat{\alpha}(z_i) + \frac{3}{4}h^2\Delta_g \hat{\alpha}(z_i)\Bigr)|\Psi\rangle +\mathcal{O}(h^3),
\end{align}
and so the linear term should not affect observables. We leave the detailed analysis of these boundary effects to future work, since obviously they do not significantly affect the accuracy of the observables computed in this work.

\begin{center}
\begin{figure}[t]
\centering
	\includegraphics[width=8.5cm]{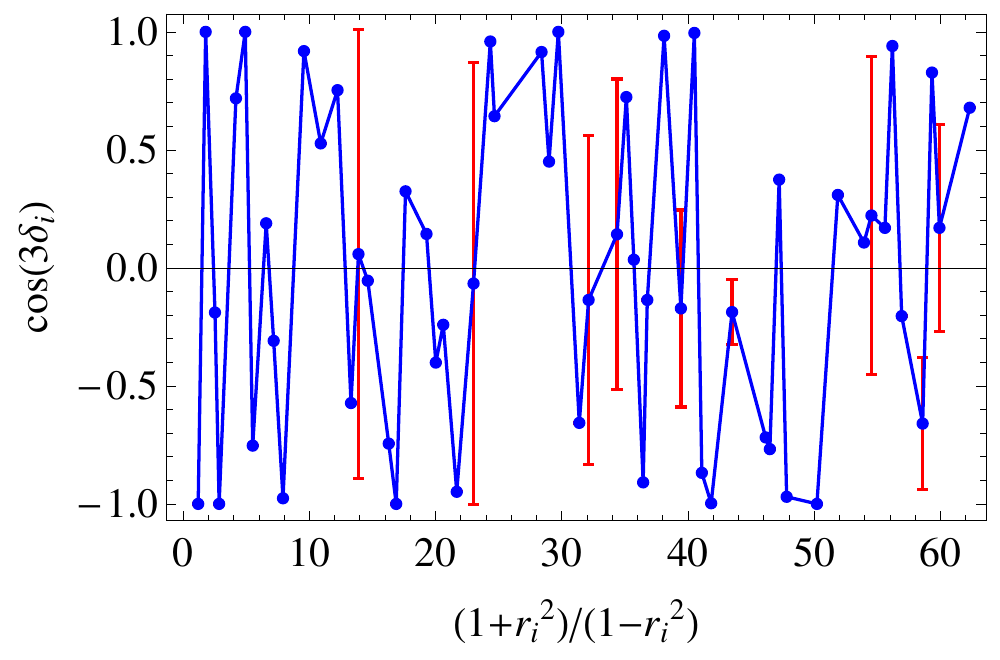}
\caption{The angle $\delta_i=\chi_i-\phi_i$ is heavily oscillating between sites with nearby radius. In most cases, $\delta_i$ only depends on the value of the radius $|z_i|=r_i$. Here we plot the averaged angle $\delta_i$ vs. site radius $r_i$ for the first 5 rings (comprising 847 sites). The red ``error bars'' indicate those cases where the mapping $\delta_i \leftrightarrow r_i$ is not unique. Still, for any given site $i$ we can unambiguously assign the value of $\delta_i$ through Eqs. (\ref{adj7})-(\ref{adj9}).}
 \label{FigScos3delta}
\end{figure}
\end{center}

\emph{Third order in $h$.} It is further possible to determine the coefficient $Q_3$ in the expansion
\begin{align}
 \label{adj28} A_{ij} f_j = 3 f(z_i)  + \frac{3}{4} h^2 \Delta_g f(z_i) + h^3Q_3  + \mathcal{O}(h^4)
\end{align}
along the same lines. We have
\begin{align}
\label{adj32}  Q_3 &= -\frac{1}{2}\Bigl[e^{3\rmi \chi_i}\mathcal{D}f(z_i)+e^{-3\rmi \chi_i}\bar{\mathcal{D}}f(z_i)\Bigr]
\end{align}
with differential operator
\begin{align}
\label{adj33} \mathcal{D} &= \partial_z^2(1-|z|^2)^3\partial_z.
\end{align}
To understand the role of $\chi_i$ in this formula, it is instructive to express $\partial_z$ in radial coordinates according to
\begin{align}
 \partial_z = \frac{e^{-\rmi \phi}}{2}\Bigl(\partial_r -\frac{\rmi}{r} \partial_\phi\Bigr).
\end{align} 
Consequently only the combination 
\begin{align}
\delta_i = \chi_i -\phi_i
\end{align}
is relevant when applying $\mathcal{D}$ in Eq. (\ref{adj23}). We find that $\delta_i$ in most cases (not all) only depends on the radius $r_i$, and is wildly fluctuating as we go from one lattice site to the other, see Fig. \ref{FigScos3delta}. This may explain why the contributions of the operator $\mathcal{D}$, although only suppressed by a power of $h$, seem to be unimportant for computing the observables we consider in this work.

\section{Finite-size scaling limit}\label{AppFS}
In this section, we perform a formal finite-size scaling limit $h\to 0$ of Eq. (\ref{adj14}) and show that the higher-order corrections vanish in this formal limit.

The lattice parameter $h$ is fixed by hyperbolic geometry for every $\{p,3\}$ lattice through $h=\tanh(d_0)$ with $d_0=d(r_0,r_0 e^{2\pi \rmi/p})$ and $r_0$ from Eq. (\ref{Eqr0}). Some examples for the lowest (and highest) values of $p$ are:

\begin{table}[h]
\begin{tabular}{|c|c|c|c|c|c|}
\hline
 $p$ & $7$ & $8$ & $9$ & $10$ & $\infty$ \\
\hline
 \ $d_0$ \ & \ 0.283128 \  &\ 0.36352 \ &\ 0.409595 \ &\ 0.439590 \ & 1 \\
\hline
 $h$ & \ 0.275798 \ &\ 0.348311 \ &\ 0.388129 \ &\ 0.413304 \ &\ $1/2$ \ \\
\hline
\end{tabular}
\end{table}

\noindent We observe that the heptagonal lattice has the smallest value of $h$. As such, the fixed value of $h$ yields a fundamental limit to the accuracy of the approximation in Eq. (\ref{adj26}). However, since $h$ is relatively small, including the $\mathcal{O}(h^3)$ correction in Eq. (\ref{adj28}) yields virtually exact results.

Remarkably, the right-hand side of Eq. (\ref{adj14}) can be evaluated for every value of $h$ and so allows for a formal finite-size scaling limit $h\to 0$. To see this, let us rewrite the equation as
\begin{align}
 \nonumber (\hat{A}_hf)_i :={}& f\Bigl(\frac{z_i-he^{\rmi \chi_i}}{1-he^{\rmi \chi_i}\bar{z}_i}\Bigr) + f\Bigl(\frac{z_i-h e^{\rmi 2\pi/3}e^{\rmi \chi_i}}{1-h e^{\rmi 2\pi/3}e^{\rmi \chi_i}\bar{z}_i}\Bigr) \\
 \label{adj34} &+ f\Bigl(\frac{z_i-h e^{\rmi 4\pi/3}e^{\rmi \chi_i}}{1-h e^{\rmi 4\pi/3}e^{\rmi \chi_i}\bar{z}_i}\Bigr),
\end{align}
where we defined a formal, $h$-dependent, operator $\hat{A}_h$. Importantly, we can now modify the value of $h$ at will. (But, for $h\neq 0.275798$, the operator loses its interpretation as the adjacency matrix of the heptagonal lattice.)

\begin{figure}[t]
\centering
\includegraphics[width=8cm]{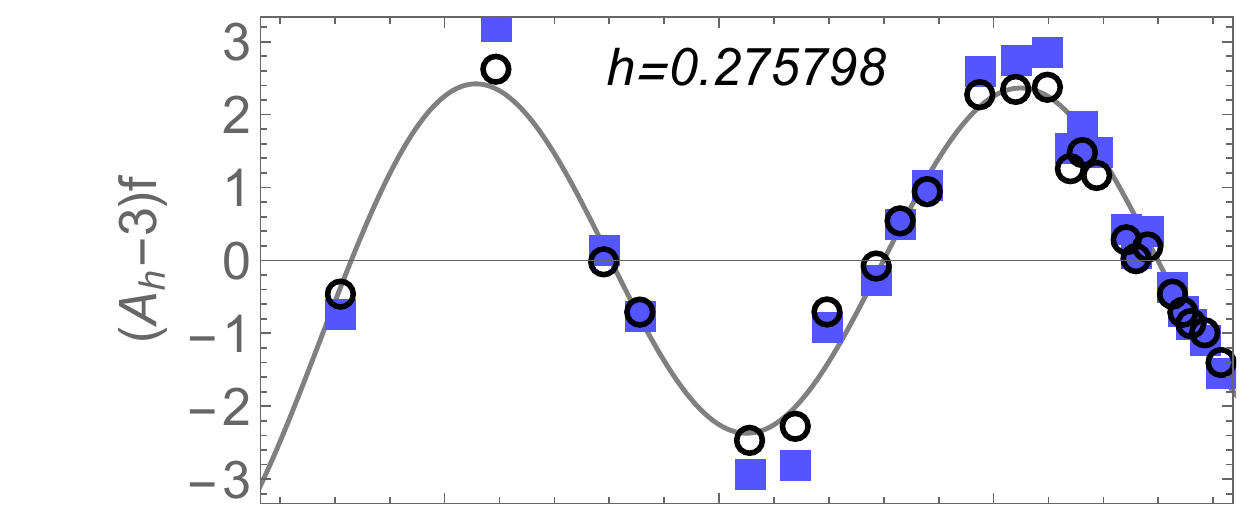}
\includegraphics[width=8cm]{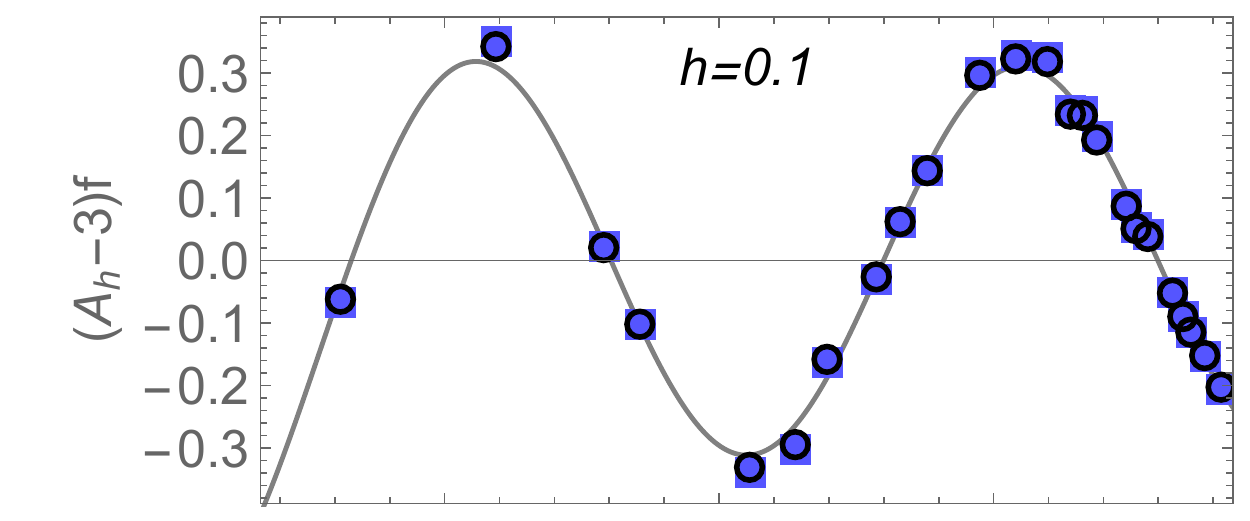}
\includegraphics[width=8cm]{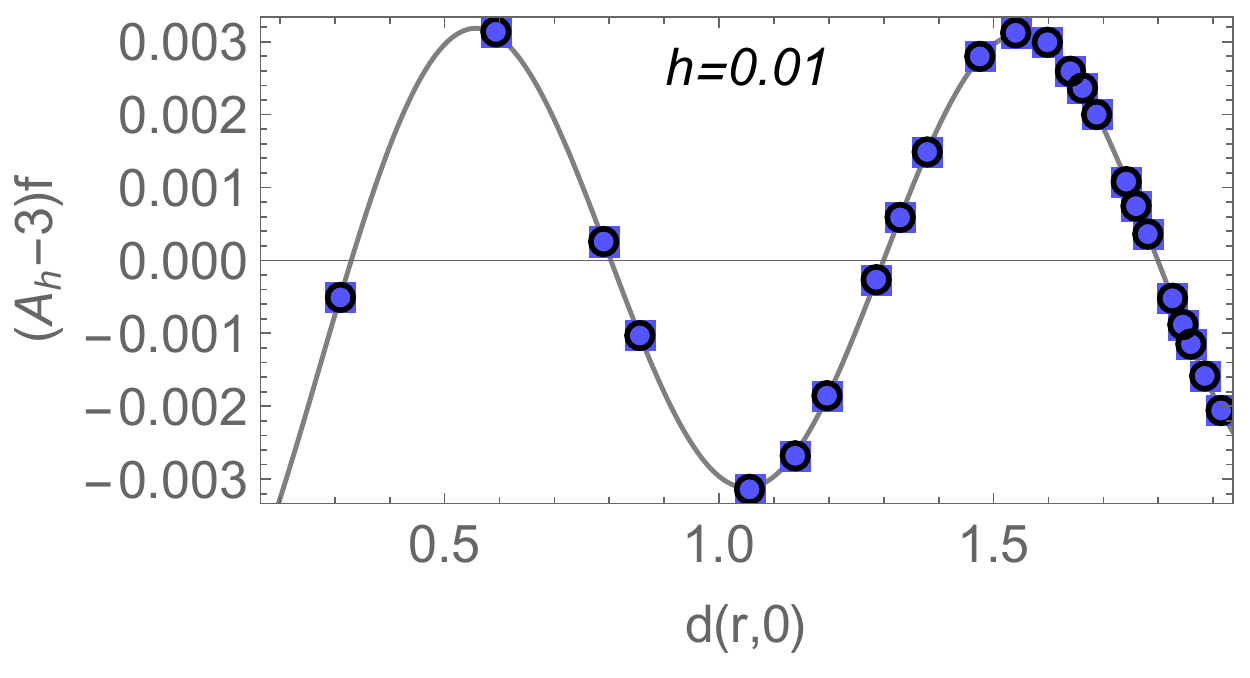}
\caption{Finite-size scaling. We define the $h$-dependent operator $\hat{A}_h$ through Eq. (\ref{adj34}), which allows us to connect the physical value of $h=0.275798$ for the heptagonal lattice to a formal $h\to 0$ limit. For the radially symmetric test function $f(r) = \cos[\pi d(r,0)]$ we show here $\hat{A}_hf$ (empty black circles), the cubic approximation $\hat{A}_h^{(3)}f$ from Eq. (\ref{adj36}) (filled blue squares), and the quadratic approximation $\hat{A}_h^{(2)}=3f+\frac{3}{4}h^2\Delta_g f$ (gray solid line).  We subtract $3f$ for better visibility of the deviations. The three figures correspond to decreasing values of $h$, with the individual values given in the plot labels. As $h\to 0$, the quadratic approximation quantitatively reproduces the full expression.}
\label{FigSh}
\end{figure}

To simplify the finite-size analysis, let us focus on radially symmetry functions $f(r)$. For most sites $i$, the expression $(\hat{A}_hf)_i$ only depends on the radius $r_i$, see Fig. \ref{FigScos3delta}. Consequently, after averaging over the radii, we can plot $\hat{A}_hf$ vs. the hyperbolic distance from the origin $d(r,0)$, see Fig. \ref{FigSh}. We compare the full expression $\hat{A}_hf$ with the quadratic and cubic continuum approximations given by
\begin{align}
\label{adj35} (\hat{A}_h^{(2)}f)_i = 3f(z_i) + \frac{3}{4}h^2\Delta_gf(z_i)
\end{align}
and
\begin{align}
 \nonumber (\hat{A}_h^{(3)}f)_i ={}& 3f(z_i) + \frac{3}{4}h^2\Delta_gf(z_i)\\
 \label{adj36} &-\frac{h^3}{2}\Bigl[e^{3\rmi \chi_i}\mathcal{D} f(z_i)+ e^{-3\rmi \chi_i}\bar{\mathcal{D}} f(z_i)\Bigr],
\end{align}
respectively. We verify that for $h\to 0$, the quadratic approximation is sufficient to approximate the full result $\hat{A}_hf$. For general $h\leq 0.275798$, $\hat{A}_h^{(2)}f$ always gives a good qualitative and overall approximation. For sizeable but small $h\sim 0.1$, the cubic approximation is sufficient to even capture the quantitative behavior of $\hat{A}_hf$, whereas for the physical value $h=0.275798$, deviations between $\hat{A}_hf$ and $\hat{A}^{(3)}_hf$ are visible, though small.

\section{Derivation of Eqs. (8) and (9)}\label{AppSum}

In this section we show how sums of the type
\begin{align}
\label{sum1} \sum_{i\in G} f(z_i)
\end{align}
with a suitable function $f(z)$ can be approximated by integrals over the Poincar\'{e} disk with a finite radius $L<1$. We assume the graph $G$ to consist of $\ell$ rings. We present two methods: Method 1 is geometric in nature and relies on determining the area of the fundamental polygon of the dual lattice, whereas Method 2 is numerical and based on fitting the growth of the number of sites with increasing radial coordinate. The outcome, Eqs. (8) and (9) in the main text, is the same.

\emph{Method 1.} The first method to approximate sums by integrals employs the fact that \cite{BookFEM,PhysRevD.98.014502,PhysRevD.95.114510,brower2019lattice}
\begin{align}
 \label{sum1b} \sum_{i\in G} \mathcal{A}_i f(z_i) \simeq \int \frac{\mbox{d}^2z}{(1-|z|^2)^2} f(z)
\end{align}
with $\mathcal{A}_i=\text{area}(\mathcal{P}_i)$, where $\mathcal{P}_i=\{z\in \mathbb{D} \ : d(z,z_i)\leq d(z,z_j)$ for all $j \neq i\}$ is the set of all points in $\mathbb{D}$ that are closer to $z_i$ in comparison to any other lattice point. Equation (\ref{sum1b}) is commonly applied for discretization of curved manifold, such as in finite-elements methods for numerical simulations. Due to the high symmetry of the tessellation $\{p,3\}$ with regular $p$-gons, the set $\mathcal{P}_i$ is the fundamental triangle of the dual lattice $\{3,p\}$, see Fig. \ref{Fig3}. Furthermore, the area $\mathcal{A}_i =\mathcal{A}_\bigtriangleup$ is independent of $i$ in this case, as all sites are equivalent with respect to automorphisms. We then arrive at Eq. (8) from the main text.

In order to compute the area $\mathcal{A}_\bigtriangleup$, we note that the area of a hyperbolic triangle with interior angles $\alpha, \beta, \gamma$ is generally given by $(\pi-\alpha-\beta-\gamma)/4$. In this case, all interior angles are $2\pi /p$. To see this, choose a dual triangle with one vertex at the origin. Clearly, the interior angle at the site at the origin is $2\pi/p$, because the outgoing geodesics are straight lines. However, since the other two vertices of the triangle can be brought to the origin by a suitable Moebius transformation, we conclude that in fact all interior angles are $2\pi/p$. We then arrive at
\begin{align}
 \label{sum1c} \mathcal{A}_\bigtriangleup = \frac{\pi}{4}\Bigl(1-\frac{6}{p}\Bigr),
\end{align}
as quoted in the main text. For $p=7$, we have $\mathcal{A}_\bigtriangleup = \pi/28$.

\begin{center}
\begin{figure}[t!]
\centering
	\includegraphics[width=8cm]{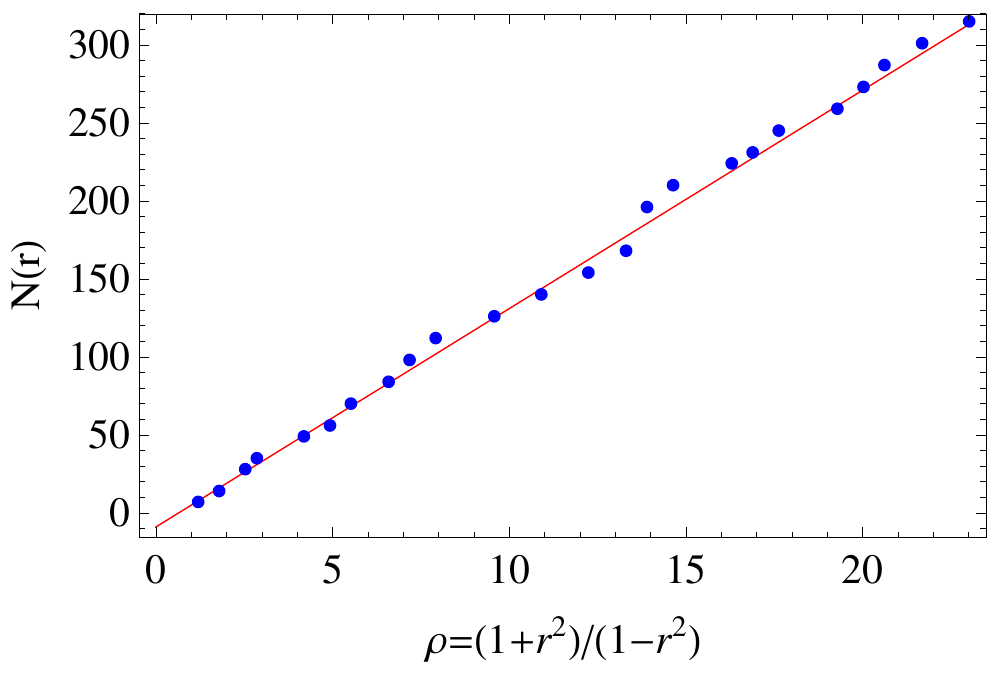}
\caption{Counting function $\mathcal{N}(r)$ for radii within the first $\ell=4$ rings versus the linear curve $14\rho+b$, with $b=-9.096$.}
 \label{FigS1}
\end{figure}
\end{center}

\emph{Method 2.} To simplify the matter let us first assume that $f(z)$ only depends on $r=|z|$. Define the counting function
\begin{align}
 \label{sum2} \mathcal{N}(r) = \sum_{i\in G}\Theta(r-r_i).
\end{align}
Further introduce the hyperbolic invariant $\rho$ via
\begin{align}
 \label{sum3} \rho =\frac{1+r^2}{1-r^2},\ \mbox{d}\rho = 4 \frac{\mbox{d}r\ r}{(1-r^2)^2}.
\end{align}
We find that $\mathcal{N}(r)$ is approximately linear in $\rho$ and given by
\begin{align}
 \label{sum4} \mathcal{N}(r) \approx 14 \rho+b,
\end{align}
with $b$ a constant, see Fig. \ref{FigS1}. Hence $\mbox{d}\mathcal{N}(r) = 14\mbox{d}\rho$. In order to approximate the finite sum by a compactly supported integral, we restrict the integration to a disk of radius $L<1$, with $L$ to be determined. We have
\begin{align}
 \nonumber \sum_{i\in G} f(r_i) &\approx \int \mbox{d}N(r)\ f(r) = 14 \int \mbox{d}\rho\ f(r)\\
 \nonumber &= 14\cdot4\int_0^{L} \frac{\mbox{d}r\ r}{(1-r^2)^2}f(r) \\
 \label{sum5} &=\frac{14\cdot 4}{2\pi} \int_{|z|\leq L} \frac{\mbox{d}^2z}{(1-|z|^2)^2} f(|z|).
\end{align}
We fix the effective radius $L$ by matching the total number of sites $N$ to the right-hand side $\frac{14\cdot4}{2\pi} \frac{\pi L^2}{1-L^2}$ for $f(r)=1$. This yields
\begin{align}
\label{sum6} L = \sqrt{\frac{N}{N+28}}.
\end{align}

In order to approximate the angular dependence for a more general function $f(z)=f(r e^{\rmi \phi})$ we can employ standard arguments from Riemann integration. Due to the sevenfold rotation invariance of the lattice we can divide the $N$ lattice sites $\{i\in G\}$ into $N/7$ ``shells'', labelled $[i]$, where each shell contains seven sites with equal radius $r_i$ and angle $\phi_i+2\pi j/7$ with $j=0,\dots,6$. Summing over these seven sites for fixed $r_i$ we have
\begin{align}
\label{sum7}  \sum_{j=0}^6 f( r_i e^{\rmi \phi_i+2\pi \rmi j/7}) \approx 7 \int_0^{2\pi} \frac{\mbox{d}\phi}{2\pi} f(r_i e^{\rmi \phi}),
\end{align}
where we approximated the sum by a Riemann integral. For the total sum over lattice sites we then have
\begin{align}
 \nonumber \sum_{i\in G} f(r_i e^{\rmi \phi_i}) &= \sum_{[i]}  \sum_{j=0}^6 f( r_i e^{\rmi \phi_i+2\pi \rmi j/7})\\
 \nonumber &\approx \int_0^{2\pi} \frac{\mbox{d}\phi}{2\pi} \Bigl( \sum_{[i]} 7 f(r_i e^{\rmi \phi})\Bigr)\\
 \nonumber &= \int_0^{2\pi} \frac{\mbox{d}\phi}{2\pi}  \sum_{i\in G}  f(r_i e^{\rmi \phi})\\
 \nonumber &\approx \int_0^{2\pi} \frac{\mbox{d}\phi}{2\pi}  \cdot 14\cdot4\int_0^{L} \frac{\mbox{d}r\ r}{(1-r^2)^2}f(r e^{\rmi \phi})\\
 \label{sum8} &= \frac{14\cdot 4}{2\pi} \int_{|z|\leq L} \frac{\mbox{d}^2z}{(1-|z|^2)^2} f(z).
\end{align}
This shows that Eq. (\ref{sum5}) generalizes to functions with nontrivial angular dependence.

\section{Eigenfunctions of hyperbolic Laplacian}\label{AppEig}

In this section we summarize the spectral properties of the hyperbolic Laplacian, i.e. the eigenvalues and eigenfunctions in different representations. 

We label the eigenfunctions of $\Delta_g$ on the Poincar\'{e} disk by a momentum parameter
\begin{align}
 \label{eig10} K= k e^{\rmi \beta}
\end{align}
with $k\geq0$ the amplitude and $\beta\in[0,2\pi)$ a phase. The corresponding eigenfunction is
\begin{align}
 \label{eig12} \psi_K(z) = \Bigl(\frac{1-|z|^2}{|1-z e^{-\rmi \beta}|^2}\Bigr)^{\frac{1}{2}(1+\rmi k)},
\end{align}
and applying $\Delta_g=4(1-|z|^2)^2\partial_z\bar{\partial}_z$ it is easy to see that the corresponding eigenvalue is
\begin{align}
 \label{eig12b} \vare_K = -(k^2+1).
\end{align}
The norm of $\psi_K$ for a disk of radius $L<1$ is
\begin{align}
 \label{eig13} ||\psi_K||^2 = \int_{|z|\leq L} \frac{\mbox{d}^2z}{(1-|z|^2)^2} |\psi_K(z)|^2 = \frac{\pi L^2}{1-L^2},
\end{align}
which is the hyperbolic volume of the disk. Note how this is analogous to the Euclidean case. Indeed, the Euclidean plane wave solutions
\begin{align}
 \label{eig14} \psi_K(z) \simeq e^{\rmi \textbf{k}\cdot \textbf{x}} = \exp\Bigl[\frac{\rmi}{2}(K \bar{z}+\bar{K}z)\Bigr]
\end{align}
are recovered for $z\ll 1$ and $k\gg 1$, which corresponds to the radius of the Poincar\'{e} disk approaching infinity and hence vanishing curvature.

Often it is advantageous to express the eigenfunction in radial coordinates $z=r e^{\rmi \phi}$. For $k\geq0$ and $m\in\mathbb{Z}$ we define the radial eigenfunctions $g_{km}(r)$ corresponding to the eigenvalue $-(k^2+1)$ such that
\begin{align}
\label{eig15} \psi_K(z) = \sum_{m=-\infty}^\infty \rmi^m g_{km}(r) e^{\rmi m (\phi-\beta)}.
\end{align} 
The ansatz is motivated by the Euclidean formula
\begin{align}
 \label{eig16} e^{\rmi \textbf{k}\cdot\textbf{x}} = \sum_{m=-\infty}^\infty \rmi^m J_m(kr) e^{\rmi m (\phi-\beta)}
\end{align}
with Bessel functions $J_m$. We have
\begin{align}
 \label{eig17} g_{km}(r) = \rmi^{-m} \int_0^{2\pi} \frac{\mbox{d}\phi}{2\pi} e^{-\rmi m\phi} \Bigl(\frac{1-r^2}{|1-r e^{\rmi \phi}|^2}\Bigr)^{\frac{1}{2}(1+\rmi k)},
\end{align}
which can be solved numerically. However, the functions $g_{km}(r)$ can be determined in closed form by making the ansatz $g_{km}(r)= \hat{g}(\frac{1+r^2}{1-r^2})$, which leads to Legendre's differential equation (\ref{green16}) for $\hat{g}$. This yields
\begin{align}
  \label{eig18} g_{km}(r) \propto P_\nu^m\Bigl(\frac{1+r^2}{1-r^2}\Bigr)
\end{align}
with $\nu=\frac{1}{2}(-1+\rmi k)$ and $P_\nu^m$ the Legendre function of the first kind. The correct prefactor ensuring Eq. (\ref{eig15}) is found to be
\begin{align}
 \nonumber g_{k,m=0}(r) &= P_{\nu}\Bigl(\frac{1+r^2}{1-r^2}\Bigr),\\
 \label{eig19} g_{k,m>0}(r) &=\frac{1}{\prod_{n=0}^{m-1}(\nu-n)} P_{\nu}^m\Bigl(\frac{1+r^2}{1-r^2}\Bigr),\\
 \nonumber  g_{k,m<0}(r) &= (-1)^m g_{k,m>0}(r).
\end{align}
In the Euclidean limit $r\ll1 $ and $k\gg1$ we recover $g_{km}(r) \simeq J_m(kr)$.

\section{Computation of lowest eigenvalues and derivation of Eq. (14)}\label{AppSpec}

In this section we give details for the determination of the ground state energy and spectral gap of (1) the Hamiltonian $H=-A$ on a graph with $\ell$ rings and (2) its continuum approximation $H^{(\rm cont)}=-3-\frac{3}{4}h^3\Delta_g$ on a disk of radius $L=\sqrt{N/(N+28)}$.

\emph{Graph Hamiltonian.} Denote the lowest two eigenvalues of $H$ by $E_0$ and $E_1>E_0$. For moderately sized $\ell \lesssim 7$ we easily find the spectrum of $A$ using matrix diagonalization. For larger $\ell\geq 8$, due to the exponential increase in the size of the matrix, we use sparse matrix techniques to determine the lowest two eigenvalues. Specifically, we employ the numerical Lanczos algorithm to estimate $E_0$ and $E_1$, and estimate the results to be reliable to about three significant digits. Note that the accuracy of the algorithm becomes worse as the spacing between $E_0$ and $E_1$ becomes smaller. Since a more precise determination of the values of $E_0$ and $E_1$ is not among the goals of the present work, we are not able to test whether the relative error between $E_{0,1}$ and $E_{0,1}^{(\rm cont)}$ decreases as $\ell \to \infty$. The results are summarized in Table \ref{SuppTab1}.

\emph{Continuous Hamiltonian.} The eigenvalues of $H^{(\rm cont)}$ read
\begin{align}
 \label{spec1} E_{nm} = -3 + \frac{3}{4}h^2 (k_{nm}^2+1),
\end{align}
where $k_{nm}$ satisfies the Dirichlet boundary condition
\begin{align}
 \label{spec2} P^m_{\nu_{nm}}\Bigl(\frac{1+L^2}{1-L^2}\Bigr)=0
\end{align}
with Legendre function of the first kind $P^m_\nu$ and $\nu_{nm}=\frac{1}{2}(-1+\rmi k_{nm})$. We label the eigenenergies by $n=1,2,3,\dots$ and $m\in\mathbb{Z}$. The ground and first excited state have azimuthal quantum numbers $m=0$ and $m=1$, respectively. In order to declutter notation in the following we define
\begin{align}
 \label{spec3} k_0&= k_{1,0},\ E_0^{(\rm cont)} =-3 + \frac{3}{4}h^2 (k_0^2+1),\\
 \label{spec4} k_1&= k_{1,1},\ E_1^{(\rm cont)} =-3 + \frac{3}{4}h^2 (k_1^2+1).
\end{align}
Equation (\ref{spec2}) is readily solved numerically and we present the lowest two eigenvalues $E_0^{(\rm cont)}$ and $E_1^{(\rm cont)}$ in Table \ref{SuppTab1}.

\renewcommand{\arraystretch}{1.6}
\begin{table*}[t]
\begin{tabular}{|c||c|c|c|c|c|c|c|c|c|c|}
\hline $\ell$ & 1 & 2 & 3 & 4 & 5 & 6 & 7 & 8 & 9 & 10\\
\hline\hline $E_0$  & $-2$ & $ -2.636$ & $-2.787$ & $-2.847$ & $-2.877$ & $-2.894$ & $-2.905$ & $-2.91$   & $-2.92$ & $-2.92$ \\
 \hline exact $E_0^{(\rm cont)}$  & $-1.500$ & $-2.570$ & $-2.770$ & $-2.842$ & $-2.876$ & $-2.895$ & $-2.906$ & $-2.914$ & $-2.920$ & $-2.924$ \\
 \hline asymptotic $E_{0}^{(\rm cont)}$ & $-2.535$ & $-2.738$ & $-2.820$ & $-2.861$ & $-2.884$ & $-2.899$ & $-2.909$ &  $-2.916$ & $-2.921$ & $-2.924$ \\
 \hline\hline $E_1$ & $-1.274$ & $-2.283$  & $-2.627$ & $-2.762$  & $-2.827$ & $-2.863$  & $-2.884$ & $-2.90$ & $-2.91$ & $-2.91$\\ 
 \hline exact $E_1^{(\rm cont)}$ & $0.620$ & $-2.085$ &  $-2.578$ & $-2.746$ & $-2.821$ & $-2.861$ &  $-2.884$ & $-2.899$ & $-2.908$ &  $-2.915$ \\
 \hline asymptotic $E_{1}^{(\rm cont)}$ & $15.58$ &  $-1.633$ &  $-2.507$ & $-2.728$ & $-2.815$ & $-2.858$ & $-2.883$ & $-2.898$ & $-2.908$ &  $-2.915$ \\ 
\hline
\end{tabular}
\caption{We compare the ground and first excited state energies for the graph and conitnuum for $\ell=1,\dots,10$. The ``asymptotic'' continuum formulas for $E_{0}^{(\rm cont)}$ and $E_{1}^{(\rm cont)}$ correspond to Eqs. (\ref{spec11}) and (\ref{spec17}), respectively. For $\ell\geq 8$ we give an estimate of $E_0$ and $E_1$ from sparse matrix methods.}
\label{SuppTab1}
\end{table*}
\renewcommand{\arraystretch}{1}

We can employ Eq. (\ref{spec2}) to compute the asymptotic behavior of $k_0$ as $L\to 1$ (or equivalently $\ell\to \infty$). For large $x\to \infty$ we have
\begin{align}
 \nonumber P_\nu(x\to \infty) &\sim \Bigl(\frac{2}{x}\Bigr)^{\frac{1}{2}(1+\rmi k)} \frac{\Gamma(-\rmi k)}{\Gamma(\frac{1}{2}(1-\rmi k))^2}\\
 \label{spec6} &+\Bigl(\frac{2}{x}\Bigr)^{\frac{1}{2}(1-\rmi k)}\frac{\Gamma(\rmi k)}{\Gamma(\frac{1}{2}(1+\rmi k))^2}
\end{align}
with Euler's $\Gamma$-function. In order to find the zeros we consider the amplitude
\begin{align}
  \label{spec7} |P_\nu(x\to \infty)|^2 &\sim \frac{2}{x}\Bigl|\frac{\Gamma(-\rmi k)}{\Gamma(\frac{1}{2}(1-\rmi k))^2}\Bigr|^2\\
 \nonumber &\times\Bigl|1+\Bigl(\frac{2}{x}\Bigr)^{-\rmi k} \frac{\Gamma(\rmi k)\Gamma(\frac{1}{2}(1-\rmi k))^2}{\Gamma(-\rmi k)\Gamma(\frac{1}{2}(1+\rmi k))^2}\Bigr|^2\\
  \nonumber &=\frac{2}{x}\frac{\cosh^2(\pi k/2)}{\pi k \sinh(\pi k)}\times\Bigl|1+\Bigl(\frac{2}{x}\Bigr)^{-\rmi k} e^{\rmi \Phi(k)}\Bigr|^2
\end{align}
with
\begin{align}
 \label{spec8} \rmi \Phi(k) = \ln \Bigl(\frac{\Gamma(\rmi k)\Gamma(\frac{1}{2}(1-\rmi k))^2}{\Gamma(-\rmi k)\Gamma(\frac{1}{2}(1+\rmi k))^2}\Bigr).
\end{align}
The first term in Eq. (\ref{spec7}) is positive. Thus the lowest zero $k_0$ for $L\to 1$ follows from
\begin{align}
 \nonumber \pi &\stackrel{!}{=}  -k_0 \ln\Bigl(\frac{2}{x}\Bigr)+\Phi(k_0)\\
 \label{spec9} &= -k_0 \ln\Bigl(\frac{2}{x}\Bigr) - \pi +4k_0\ln2 + \mathcal{O}(k^3)\\
 \nonumber &= -\pi + k_0 \ln(8x) + \mathcal{O}(k^3)
\end{align}
for $x=\frac{1+L^2}{1-L^2}\to \infty $. Consequently,
\begin{align}
\label{spec10} k_0 &\sim \frac{2\pi}{\ln(8 \frac{1+L^2}{1-L^2})} \simeq \frac{\pi}{\ln (2\vphi^\ell)}.
\end{align}
We inserted $N\sim 7 \vphi^{2\ell}$ with golden ratio $\vphi=(1+\sqrt{5})/2$ such that $\frac{1+L^2}{1-L^2} \sim \frac{1}{2}\vphi^{2\ell}$. The ground state energy for large $\ell$ follows as
\begin{align}
  \label{spec11} E_0^{(\rm cont)} \sim -3 + \frac{3}{4}h^2 + \frac{3\pi^2h^2}{4} \frac{1}{[\ln \vphi\cdot \ell+\ln2]^2}
\end{align}

The first excited state energy of $H^{(\rm cont)}$ can be determined in a fully analogous way from
\begin{align}
 \label{spec12} P_{\nu_1}^{1}\Bigl(\frac{1+L^2}{1-L^2}\Bigr)=0,\ \nu_1=\frac{1}{2}(-1+\rmi k_1).
\end{align}
We have
\begin{align}
 \label{spec13} |P_{\nu}^1(x\to \infty)|^2 \propto \Bigl| 1+\Bigl(\frac{2}{x}\Bigr)^{-\rmi k} e^{\rmi \Phi_1(k)}\Bigr|^2
\end{align}
with
\begin{align}
 \label{spec14} \rmi \Phi_1(k) = \ln\Bigl( \frac{\Gamma(\rmi k)\Gamma(\frac{1}{2}(1-\rmi k))\Gamma(\frac{1}{2}(-1-\rmi k))}{\Gamma(-\rmi k)\Gamma(\frac{1}{2}(1+\rmi k))\Gamma(\frac{1}{2}(-1+\rmi k))}\Bigr).
\end{align}
Hence $k_1$ for $x=\frac{1+L^2}{1-L^2}\to \infty$ is found from
\begin{align}
 \nonumber \pi &\stackrel{!}{=} -k_1\ln\Bigl(\frac{2}{x}\Bigr) + \Phi_1(k_1)\\
 \label{spec15} &=  -k_1\ln\Bigl(\frac{2}{x}\Bigr) -\pi + 2[2\ln2-1]k_1+\mathcal{O}(k_1^3)\\
 \nonumber &= -\pi -k_1\ln\Bigl(\frac{8x}{e^2}\Bigr)+\mathcal{O}(k_1^3).
\end{align}
This implies
\begin{align}
 \label{spec16} k_1 \sim \frac{2\pi}{\ln(\frac{8}{e^2}\frac{1+L^2}{1-L^2})} \simeq \frac{\pi}{\ln(\frac{2}{e}\vphi^\ell)}
\end{align}
and
\begin{align}
 \label{spec17} E_1^{(\rm cont)} \sim  -3 + \frac{3}{4}h^2+\frac{3\pi^2h^2}{4} \frac{1}{[\ln \vphi\cdot\ell  + \ln2-1]^2}.
\end{align}

\section{Derivation of Eq. (16)}\label{AppPath}

In this section we derive the relation between the graph green function $G_{ij}(\omega)$ and the continuum Green function $G(z,z',\lambda,L)$ using an auxiliary field path integral representation.

\vspace{0.5cm}
We first recall some path integral identities \cite{ZinnJustinBook}. If $\theta_i$ is a discrete real variable and $M_{ij}$ a real and symmetric matrix then
\begin{align}
\label{path1} \langle \theta_i \theta_j\rangle &:= \frac{\int\mbox{D}\theta\ \theta_i \theta_j\ e^{-\frac{1}{2}\theta_k M_{kl}\theta_l}}{\int\mbox{D}\theta\  e^{-\frac{1}{2}\theta_k M_{kl}\theta_l}} = (M^{-1})_{ij}
\end{align}
with $\mbox{D}\theta=\prod_i \mbox{d}\theta_i$. Furthermore, if $\theta(x)$ is a real field and $\hat{D}$ a differential operator, then $\mbox{D}\theta=\prod_x \mbox{d}\theta(x)$ and
\begin{align}
 \label{path2} \langle \theta(x)\theta(y) \rangle := \frac{\int\mbox{D}\theta\ \theta(x)\theta(y)  e^{-\frac{1}{2}\int d^dr \theta \hat{D}\theta}}{\int\mbox{D}\theta\  e^{-\frac{1}{2}\int d^dr \theta \hat{D}\theta}} = - G(x,y),
\end{align}
where $G(x,y)$ is the Green function of $\hat{D}$ according to
\begin{align}
 \label{path3} \hat{D} G(x,y) = -\delta^{(d)}(x-y).
\end{align}

This can be applied to match the Green function on the graph $G_{ij}$ to the continuum Green function $G(z,z')$ evaluated on the graph sites. The graph Green function is given by
\begin{align}
 \label{path4}  G_{ij}(\omega,\ell) = \Bigl(\frac{1}{H-\omega\mathbb{1}}\Bigr)_{ij}  = \sum_{n=1}^{N} \frac{\psi_n(z_i)\psi_n^*(z_j)}{\vare_n-\omega}
\end{align}
with $H=-A$ and $H\psi_n =\vare_n \psi_n$. Introduce the auxiliary real field $\theta_i\to\theta(z_i)$ and approximate $H \approx -3-\frac{3}{4}h^2\Delta_g$ and $\sum_i \approx \frac{28}{\pi} \int_{|z|\leq L}\frac{d^2z}{(1-|z|^2)^2}$ to find
\begin{widetext}
\begin{align}
 \nonumber G_{ij}(\omega,\ell) &= \frac{ \int\mbox{D}\theta\ \theta_i \theta_j \exp\Bigl[-\frac{1}{2}\sum_{i,j} \theta_i (H-\omega \mathbb{1})_{ij}\theta_j\Bigr]}{ \int \mbox{D}\theta\  \exp\Bigl[-\frac{1}{2}\sum_{i,j} \theta_i (H-\omega\mathbb{1})_{ij}\theta_j\Bigr]}\\
 \nonumber &= \frac{ \int\mbox{D}\theta\ \theta_i \theta_j \exp\Bigl[-\frac{1}{2}\sum_{i,j} \theta_i [(3\mathbb{1}-A)-(\omega+3)\mathbb{1}]_{ij}\theta_j\Bigr]}{ \int \mbox{D}\theta\  \exp\Bigl[-\frac{1}{2}\sum_{i,j} \theta_i [(3\mathbb{1}-A)-(\omega+3)\mathbb{1}]_{ij}\theta_j\Bigr]}\\
 \label{path5} &\approx \frac{ \int\mbox{D}\theta\ \theta(z_i) \theta(z_j) \exp\Bigl[-\frac{1}{2}\frac{28}{\pi} \int_{|z|\leq L} \frac{d^2z}{(1-|z|^2)^2}\ \theta(z) [-\frac{3h^2}{4}\Delta_g-(\omega+3)]\theta(z)\Bigr]}{ \int\mbox{D}\theta\  \exp\Bigl[-\frac{1}{2}\frac{28}{\pi} \int_{|z|\leq L} \frac{d^2z}{(1-|z|^2)^2}\ \theta(z) [-\frac{3h^2}{4}\Delta_g-(\omega+3)]\theta(z)\Bigr]}\\
 \nonumber &= \frac{ \int\mbox{D}\theta\ \theta(z_i) \theta(z_j) \exp\Bigl[-\frac{1}{2}C' \int_{|z|\leq L} \frac{d^2z}{(1-|z|^2)^2}\ \theta(z) [-\Delta_g-\lambda]\theta(z)\Bigr]}{ \int\mbox{D}\theta\  \exp\Bigl[-\frac{1}{2}C' \int_{|z|\leq L} \frac{d^2z}{(1-|z|^2)^2}\ \theta(z) [-\Delta_g-\lambda]\theta(z)\Bigr]}\\
\nonumber &= \frac{1}{C'} G(z_i,z_j,\lambda,L)
\end{align}
\end{widetext}
with
\begin{align}
 \label{path6}  C' = \frac{3 h^2}{4}\frac{28}{\pi}= \frac{21 h^2}{\pi} = 0.508,\ \lambda = \frac{4(\omega+3)}{3h^2}.
\end{align}
Here $G(z,z',\lambda,L)$ is the hyperbolic Green function of $(\Delta_g +\lambda)$ on a disk of radius $L$, i.e.
\begin{align}
\label{path7}  \frac{1}{(1-|z|^2)^2}(\Delta_g+\lambda) G(z,z',\lambda,L) = -\delta^{(2)}(z-z').
\end{align}
We conclude
\begin{align}
 \label{path8} G_{ij}(\omega,\ell) = \frac{\pi}{21h^2}\ G\Bigl(z_i,z_j,\lambda=\frac{4(\omega+3)}{3h^2},L\Bigr).
\end{align}

\section{Continuum Green function (Computation)}\label{AppGreen}

In this section, we compute Green's functions for the hyperbolic Laplacian $\Delta_g$ and hyperbolic Helmholtz operator $\lambda+\Delta_g$ (with $\
\lambda\in\mathbb{C}$) on the disk of radius $L\leq 1$ with Dirichlet boundary conditions. A self-contained summary of the relevant formulas is given in the next section.

\emph{Definition.} The Green function is defined by
\begin{align}
\label{green1} (\lambda+\Delta_g) G(z,z',\lambda,L) = -(1-|z|^2)^2\delta^{(2)}(z-z')
\end{align}
with $\Delta_g=(1-|z|^2)^24\partial_z\bar{\partial}_z$ acting on $z$ and Dirichlet boundary conditions such that $G(z,z',\lambda,L)=0$ for $|z|=L$ or $|z'|=L$. Here $(1-|z|^2)^2\delta^{(2)}(z-z')$ is the $\delta$-function with respect to the hyperbolic volume measure. For an arbitrary continuous function $f:\mathbb{D} \to \mathbb{C}$ it is defined by
\begin{align}
 \label{green2} \int_{\mathbb{D}} \frac{\mbox{d}^2z}{(1-|z|^2)^2}(1-|z|^2)^2 \delta^{(2)}(z-z') f(z) = f(z').
\end{align}
In Cartesian coordinates such that $z=x+\rmi y$ we have $\mbox{d}^2z=\mbox{d}x\ \mbox{d}y$ and $\delta^{(2)}(z-z')=\delta(x-x')\delta(y-y')$.

\emph{Spectral representation.} It is always possible to give a closed expression for the Green function in terms of the spectral decomposition of the operator. For this write the eigenfunctions of the hyperbolic Laplacian in radial coordinates by $g_{km}(r) e^{\rmi m \phi}$ with $g_{km}$ from Eq. (\ref{eig19}). The normalized eigenfunctions of $\Delta_g$ on the disk of radius $L<1$ are given
\begin{align}
 \label{green2d} \psi_{nm}(z) = \frac{g_{k_{nm}m}(r) e^{\rmi m \phi}}{||g_{k_{nm}m}||},
\end{align} 
where $k_{nm}$ solves $g_{k_{nm}m}(L)=0$ and the norm is
\begin{align}
\label{green2e} ||g_{km}||^2 = 2\pi \int_0^L \frac{\mbox{d}r\ r}{(1-r^2)^2}|g_{km}(r)|^2.
\end{align}
The Green function can then be written in spectral representation as
\begin{align}
 \label{green2f} G(z,z',\lambda,L) = \sum_{m=-\infty}^\infty \sum_{n} \frac{\psi_{nm}(z)\psi_{nm}^*(z')}{-\lambda+k_{nm}^2+1}.
\end{align}
Equation (\ref{green1}) follows from the completeness of the eigenfunctions and the Dirichlet boundary condition is satisfied due to $\psi_{nm}(z)=0$ for $|z|=L$. However, the spectral representation is not the most useful form of the Green function since it requires to determine the discrete momenta $k_{nm}$ and subsequently to perform the double-sum numerically. Thus we derive a few complementary expressions in the following.

\emph{Hyperbolic Laplacian.} First consider the case $\lambda=0$, where Eq. (\ref{green1}) reduces to
\begin{align}
 \label{green3} \Delta_g G(z,z,',0,L) = -(1-|z|^2)^2 \delta^{(2)}(z-z').
\end{align}
We divide by $(1-|z|^2)^2$ and observe that $G(z,z',0,L)$ coincides with the Green function of the ordinary Laplacian $\Delta$ on a disk of radius $L<1$. The latter is given by
\begin{align}
 \label{green4} G(z,z',0,L) = -\frac{1}{4\pi}\ln|z-z'|^2 + \delta G(z,z',0,L),
\end{align}
where the first term is the fundamental solution, while the second term is a harmonic function ensuring Dirichlet boundary conditions. We construct $\delta G$ from a mirror charge outside the disk, whose location is obtained from inversion on the circle, $z' \to L^2/\bar{z}'$. Hence
\begin{align}
 \label{green5} \delta G(z,z',0,L) = \frac{1}{4\pi}\ln\Bigl|z-\frac{L^2}{\bar{z}'}\Bigr|^2 +\text{const}
\end{align}
with a suitably chosen constant. We arrive at
\begin{align}
 \label{green6} G(z,z',0,L) = -\frac{1}{4\pi} \ln\Bigl|\frac{L(z-z')}{L^2-z\bar{z}'}\Bigr|^2.
\end{align}
For $z'=0$ we have
\begin{align}
 \label{green7} G(r,0,0,L) = -\frac{1}{2\pi}\ln\Bigl(\frac{r}{L}\Bigr) = -\frac{1}{2\pi}\ln\Bigl(\frac{\tanh d(r,0)}{L}\Bigr)
\end{align}
with $d(r,0)$ the distance from the origin.

\emph{Hyperbolic Helmholtz operator.} Now consider the case of arbitrary $\lambda\in\mathbb{C}$. We construct the Green function by reducing the problem to a one-dimensional Sturm--Liouville problem. We refer to Appendix C of Ref. \cite{PhysRevA.94.011602} for a detailed discussion of the procedure. Write
\begin{align}
 \label{green14b} G(z,z',\lambda,L) = \mathcal{G}_0(z,z',\lambda) - \delta \mathcal{G}(z,z',\lambda,L),
\end{align}
where $\mathcal{G}_0$ is the fundamental solution and $\delta \mathcal{G}$ is a harmonic function to ensure Dirichlet boundary conditions. To construct these functions, we first solve, for $z\neq 0$, the equation
\begin{align}
 \label{green14} (\lambda+\Delta_g)f(z)=0
\end{align}
through an ansatz
\begin{align}
 \label{green15} f(z)= \sum_{m=-\infty}^\infty f_{\lambda m}(\rho)e^{\rmi m \phi} 
\end{align}
with $\rho = \frac{1+r^2}{1-r^2}$. Then $f_{\lambda m}$ satisfies Legendre's differential equation
\begin{align}
 \label{green16} (1-\rho^2)f_{\lambda m}''-2\rho f_{\lambda m}' +\Bigl(\nu(\nu+1)-\frac{m^2}{1-\rho^2}\Bigr)f_{\lambda m}=0
\end{align}
with $\nu=\frac{1}{2}(-1+\rmi\sqrt{\lambda -1})$ or $\frac{\lambda}{4}=-\nu(\nu+1)$. The two linearly independent solutions are
\begin{align}
 \label{green17} u_m(\rho) &=  Q_\nu^{|m|}(\rho),\\
 \label{green18} v_m(\rho) &= P_\nu^{|m|}(\rho),
\end{align}
which are the Legendre functions of the second/first kind, being singular/regular at $\rho=1$. Without loss of generality we assume $m\geq 0$ in the following, otherwise replace $m\to |m|$.

Introduce the Sturm--Liouville operator
\begin{align}
 \label{green19}  L_m = -4\Bigl( \frac{\mbox{d}}{\mbox{d}\rho}\Bigl[p(\rho)\frac{\mbox{d}}{\mbox{d}\rho}\Bigr]+q_m(\rho)\Bigr)
\end{align}
with
\begin{align}
 \label{green20} p(\rho) &= (1-\rho^2),\ q_m(\rho) = \nu(\nu+1)-\frac{m^2}{1-\rho^2}.
\end{align}
We have
\begin{align}
 \label{green21} (\lambda+\Delta_g) f(z) =  \sum_{m=-\infty}^\infty e^{\rmi m \phi} L_m f_{\lambda m}(\rho).
\end{align}
The fundamental solution of $L_m$ is defined through
\begin{align}
 \label{green22} L_m G_m(\rho,\rho') = -4\delta(\rho-\rho'),
\end{align}
with the factor of $4$ for later convenience. It is given by 
\begin{align}
 \nonumber G_m(\rho,\rho') &= C_m u_m(\rho_{\rm max})v_m(\rho_{\rm min})\\
 \label{green23} &= C_m\Bigl[ u_m(\rho)v_m(\rho')\Theta(\rho-\rho') \\
 \nonumber &+ u_m(\rho')v_m(\rho)\Theta(\rho'-\rho)\Bigr],
\end{align}
where $\rho_{\rm max}$ ($\rho_{\rm min}$) is the maximum (minimum) of $\rho$ and $\rho'$, and $C_m$ a constant to be determined. To verify Eq. (\ref{green22}), use $L_m u_m=0$ and $L_mv_m=0$ and the definition of $L_m$ to find
\begin{align}
 \label{green24} L_m G_m(\rho,\rho') &=  4C_m\kappa_m(\rho)\delta(\rho-\rho'),
\end{align}
where
\begin{align}
 \label{green25} \kappa_m(\rho) &= p(\rho)\Bigl[u_m'(\rho)v_m(\rho)-u_m(\rho)v_m'(\rho)\Bigr]\\
 &= -\frac{4^m \Gamma(\frac{\nu+m+2}{2})\Gamma(\frac{\nu+m+1}{2})}{\Gamma(\frac{\nu-m+2}{2})\Gamma(\frac{\nu-m+1}{2})}
\end{align}
is constant and follows from the Wronksian of Legendre's functions \cite{AbrStegun}. Thus we have to choose $C_m = -1/\kappa_m$ and find
\begin{align}
 \label{green25b } C_m &= \frac{\Gamma(\frac{\nu-m+2}{2})\Gamma(\frac{\nu-m+1}{2})}{4^m \Gamma(\frac{\nu+m+2}{2})\Gamma(\frac{\nu+m+1}{2})}
\end{align}
or
\begin{align}
 \label{green26} C_0 &= 1,\\
 \label{green26b}  C_{m\neq 0} &= \frac{(-1)^{|m|}}{\prod_{n=0}^{|m|-1}[(n+\frac{1}{2})^2+\frac{1}{4}(\lambda-1)]}.
\end{align}

We conclude that the fundamental solution of $\lambda+\Delta_g$ is given by
\begin{align}
 \label{green27} \mathcal{G}_0(z,z',\lambda) = \frac{1}{2\pi} \sum_{m=-\infty}^\infty e^{\rmi m(\phi-\phi')} G_m(\rho,\rho').
\end{align}
Indeed, using $\sum_m e^{\rmi m(\phi-\phi')} =2\pi \delta(\phi-\phi')$ and
\begin{align}
 \nonumber (1-|z|^2)^2\delta^{(2)}(z-z') &= \frac{(1-r^2)^2}{r} \delta(r-r')\delta(\phi-\phi')\\
 \label{green28} &= 4 \delta(\rho-\rho') \delta(\phi-\phi')
\end{align}
we verify
\begin{align}
 \nonumber (\lambda+\Delta_g) \mathcal{G}_0(z,z',\lambda) &=  \frac{1}{2\pi} \sum_{m=-\infty}^\infty e^{\rmi m(\phi-\phi')} L_m G_m(\rho,\rho')\\
 \nonumber &= -\frac{1}{2\pi}\sum_{m=-\infty}^\infty e^{\rmi m (\phi-\phi')} 4\delta(\rho-\rho')\\
 \label{green29}  &= - (1-|z|^2)^2\delta^{(2)}(z-z').
\end{align}

Equation (\ref{green27}) for the fundamental solution $\mathcal{G}_0$ allows to easily construct the correction $\delta \mathcal{G}$ such that the total Green function $G=\mathcal{G}_0-\delta\mathcal{G}$ satisfies Dirichlet boundary conditions: If either $|z|=L$ or $|z'|=L$, then $\rho_{\rm max}=\frac{1+L^2}{1-L^2}$. Consequently, we choose
\begin{align}
 \nonumber &\delta\mathcal{G}(z,z',\lambda,L) \\
 \label{green31} &= \frac{1}{2\pi} \sum_{m=-\infty}^\infty e^{\rmi m (\phi-\phi')}C_m  v_m(\rho)v_m(\rho')\frac{u_m(\frac{1+L^2}{1-L^2})}{v_m(\frac{1+L^2}{1-L^2})}.
\end{align}
This is a harmonic function and satisfies $\mathcal{G}_0=\delta\mathcal{G}$ whenever $|z|=L$ or $|z'|=L$. As an illustrative example consider the central correlation function for $z'=0$. We have
\begin{align}
 \label{green32} G(z,0,\lambda,L) &= \frac{1}{2\pi}\Bigl[Q_\nu\Bigl(\frac{1+r^2}{1-r^2}\Bigr) -\frac{Q_\nu(\frac{1+L^2}{1-L^2})}{P_\nu(\frac{1+L^2}{1-L^2})} P_\nu\Bigl(\frac{1+r^2}{1-r^2}\Bigr)\Bigr],
\end{align}
clearly vanishing for $r=L$.

The Green function for $L=1$ can be given in closed form, because it can only depend on the hyperbolic invariant $d(z,z')$. Making the ansatz $\mathcal{G}(z,z',\lambda,1)=F(y)$ with 
\begin{align}
 y= \cosh(2d(z,z')) = 1+\frac{2|z-z'|^2}{(1-|z|^2)(1-|z'|^2)},
\end{align}
we find for $z\neq z'$ (or $y>1$) that 
\begin{align}
 \nonumber 0 &\stackrel{!}{=} (\lambda+\Delta_g)F(y) \\
 &= -2\Bigl[(1-y^2)F''(y)-2y F'(y)-\frac{\lambda}{4} F(y) \Bigr],
\end{align}
which again is Legendre's differential equation. The singular contribution gives the fundamental solution, because
\begin{align}
 Q_\nu (y\to 1) \sim -\ln \Bigl(\frac{1}{2}\text{arcosh} (y)\Bigr)+\text{const},
\end{align}
and the regular solution is the harmonic correction to ensure the Dirichlet boundary condition. Hence
\begin{align}
 \mathcal{G}(z,z',\lambda,1) = \frac{1}{2\pi} \Bigl[ Q_\nu(y) - \mathcal{C}\cdot P_\nu(y)\Bigr]
\end{align}
with
\begin{align}
 \mathcal{C} = \lim_{y\to \infty} \frac{Q_\nu(y)}{P_\nu(y)}.
\end{align}
Importantly, since the fundamental solution does not depend on $L$, it is \emph{always} given by\cite{marklof2012ii}
\begin{align}
 \label{green30}  \mathcal{G}_0(z,z',\lambda) = \frac{1}{2\pi} Q_{\nu} \Bigl( 1+\frac{2|z-z'|^2}{(1-|z|^2)(1-|z'|^2)}\Bigr).
\end{align}
To compute the constant $\mathcal{C}$, the parameter $\lambda$ needs to be restricted to $\lambda\in \mathbb{C}\backslash[1,\infty)$, since the spectrum $k^2+1$ of $-\Delta_g$ on the infinite disk is in the interval $[1,\infty)$. For real $\lambda<1$ this implies $\nu<-1/2$. We expand $P_\nu(y)$ and $Q_\nu(y)$ for large $y$ and arrive at
\begin{align}
 \nonumber \mathcal{C} = \frac{2^\nu\rmi^{-\nu}\pi^{3/2}\Gamma(-\nu)}{\Gamma(1+\frac{\nu}{2})\Gamma(\frac{1+\nu}{2})}\Biggl[ &-\cos\Bigl(\frac{\pi \nu}{2}\Bigr) \frac{\Gamma(1+\frac{\nu}{2})^2}{\Gamma(\frac{1-\nu}{2})^2}\\
 &+\rmi \sin\Bigl(\frac{\pi \nu}{2}\Bigr) \frac{\Gamma(\frac{1+\nu}{2})^2}{\Gamma(-\frac{\nu}{2})^2}\Biggr].
\end{align}

\section{Continuum Green function (Summary)}\label{AppGreenSum}

We summarize the expressions for the Green function $G(z,z',\lambda,L)$ derived in the previous section. We write
\begin{align}
 \label{green33} G(z,z',\lambda,L) = \mathcal{G}_0(z,z',\lambda)-\delta \mathcal{G}(z,z',\lambda,L),
\end{align}
where the first term is given by 
\begin{align}
 \label{green34} \mathcal{G}_0(z,z',\lambda) = \frac{1}{2\pi} Q_{\nu} \Bigl( 1+\frac{2|z-z'|^2}{(1-|z|^2)(1-|z'|^2)}\Bigr)
\end{align}
with $Q_\nu=Q_\nu^{m=0}$ the Legendre function of the second kind and
\begin{align}
 \label{green35} \nu = \frac{1}{2}\Bigl(-1+\rmi \sqrt{\lambda-1}\Bigr).
\end{align}
The second term reads
\begin{align}
 \nonumber &\delta \mathcal{G}(z,z',\lambda,L) = \frac{1}{2\pi} P_\nu(\rho)P_\nu(\rho')\frac{Q_\nu(\frac{1+L^2}{1-L^2})}{P_\nu(\frac{1+L^2}{1-L^2})}\\
 \label{green36} &+\frac{1}{\pi}\sum_{m=1}^\infty C_m\cos[m(\phi-\phi')] P_\nu^m(\rho)P_\nu^m(\rho')\frac{Q_\nu^m(\frac{1+L^2}{1-L^2})}{P_\nu^m(\frac{1+L^2}{1-L^2})}.
\end{align}
with $P_\nu^m$ and $Q_\nu^m$ the Legendre function of the first and second kind, hyperbolic invariant
\begin{align}
 \label{green37} \rho=\frac{1+r^2}{1-r^2},
\end{align}
and
\begin{align}
 \label{green38} C_m = \frac{\Gamma(\frac{\nu-m+2}{2})\Gamma(\frac{\nu-m+1}{2})}{4^m \Gamma(\frac{\nu+m+2}{2})\Gamma(\frac{\nu+m+1}{2})}.
\end{align}
In practice, it is sufficient to limit the sum over $m$ to the first few (typically ten or less) terms. For $L=1$ we have
\begin{align}
 \delta\mathcal{G}(z,z',\lambda,1) = \frac{\mathcal{C}}{2\pi} \  P_\nu\Bigl( 1+\frac{2|z-z'|^2}{(1-|z|^2)(1-|z'|^2)}\Bigr)
\end{align}
with $P_\nu=P_\nu^{m=0}$ and 
\begin{align}
 \nonumber \mathcal{C} = \frac{2^\nu\rmi^{-\nu}\pi^{3/2}\Gamma(-\nu)}{\Gamma(1+\frac{\nu}{2})\Gamma(\frac{1+\nu}{2})}\Biggl[ &-\cos\Bigl(\frac{\pi \nu}{2}\Bigr) \frac{\Gamma(1+\frac{\nu}{2})^2}{\Gamma(\frac{1-\nu}{2})^2}\\
 &+\rmi \sin\Bigl(\frac{\pi \nu}{2}\Bigr) \frac{\Gamma(\frac{1+\nu}{2})^2}{\Gamma(-\frac{\nu}{2})^2}\Biggr].
\end{align}

\end{appendix}

\bibliography{refs_hyp}

\end{document}